\documentclass[12pt]{article}
\usepackage{amsmath}
\usepackage{graphicx}%
\usepackage{amsfonts}%
\usepackage{amssymb}
\usepackage{enumerate}
\usepackage{natbib}
\usepackage{url} 
\usepackage{color}
\usepackage{multicol}
\usepackage{multirow}
\usepackage{booktabs}
\usepackage{makecell}
\usepackage{tikz}
\usetikzlibrary{positioning, arrows.meta}

\usepackage{float}
\restylefloat{figure}

\newcommand{\blind}{1}

\newcommand{\bbeta}{ \mbox{\boldmath $ \beta $} }

\newcommand{\balpha}{ \mbox{\boldmath $ \alpha $} }

\newcommand{\btheta}{ \mbox{\boldmath $ \theta $} }

\newcommand{\bs}{\textbf{s}}

\begin{document}

\def\spacingset#1{\renewcommand{\baselinestretch}%
{#1}\small\normalsize} \spacingset{1}

\if1\blind
{
  \title{\bf Joint Spatiotemporal Modeling of Zooplankton and Whale Populations in a Dynamic Marine Environment}
  \author{Bokgyeong Kang, Erin M. Schliep, Alan E. Gelfand,\\ Christopher W. Clark, Christine A. Hudak, Charles A. Mayo,\\ Ryan Schosberg, Tina M. Yack, and Robert S. Schick
  }
  \maketitle
} \fi

\if0\blind
{
  \bigskip
  \bigskip
  \bigskip
  \begin{center}
    {\LARGE\bf Joint Spatiotemporal Modeling of Zooplankton and Whale Populations in a Dynamic Marine Environment}
\end{center}
  \medskip
} \fi

\begin{abstract}
North Atlantic right whales are an endangered species; their entire population numbers approximately 372 individuals, and they are subject to major anthropogenic threats. They feed on zooplankton species whose distribution shifts in a dynamic and warming oceanic environment. Because right whales in turn follow their shifting food resource, it is necessary to jointly study the distribution of whales and their prey. The innovative joint species distribution modeling (JSDM) contribution here is different from anything in the large JDSM literature, reflecting the processes and data we have to work with. Specifically, our JSDM supplies a geostatistical model for expected amount of zooplankton collected at a site.  We require a point pattern model for the intensity of right whale abundance.
The two process models are joined through a latent conditional-marginal specification. Further, each species has two data sources to inform their respective distributions and these sources require novel data fusion.  What emerges is a complex multi-level model. Through simulation we demonstrate the ability of our joint specification to identify model unknowns and learn better about the species distributions than modeling them individually. We then apply our modeling to real data from Cape Cod Bay, Massachusetts in the U.S.
\end{abstract}

\noindent%
\vfill

\spacingset{1.5} 


\section{Introduction}
\label{sec:intro}

North Atlantic right whales (NARW) are an endangered species, whose habitat is both industrialized \citep{krausUrbanWhaleNorth2007} and changing \citep{recordRapidClimateDrivenCirculation2019,meyer-gutbrodRedefiningNorthAtlantic2023}. Their entire population numbers approximately 372 individuals \citep{lindedn2024}, and major anthropogenic impacts include ship-strikes \citep{kelleyAssessingLethalityShip2021}, entanglement with fishing gear \citep{knowltonFishingGearEntanglement2022}, and climate change \citep{recordRapidClimateDrivenCirculation2019}. To minimize anthropogenic impact and stop the continued population decline it is critical to infer throughout their habitat range both how many individuals there are and where they are located at any given time. 

Though climate change has impacted habitat use patterns \citep{meyer-gutbrodRedefiningNorthAtlantic2023}, for decades the canonical distribution pattern has included a calving and nursing ground off the southeastern United States, followed by spring and summertime foraging in areas like Cape Cod Bay, MA, the Great South Channel, and the Bay of Fundy and Roseway Basin\citep{winnDistributionalBiologyRight1986}. In these foraging areas, right whales feed on various copepod species \citep{sorochanAvailabilitySupplyAggregation2021}. 
Distributional shifts for copepod species as a response to warming of ocean waters \citep{grieveProjectingEffectsClimate2017} has impacted the habitat use of right whales \citep{recordRapidClimateDrivenCirculation2019,meyer-gutbrodRedefiningNorthAtlantic2023}, often with catastrophic results \citep{daoustIncidentReportNorth2017}. Thus, it is necessary to concurrently study the distribution of NARW and their prey, because their habitats are changing spatially and temporally and, due to this change, unexpected arrivals of NARW in places like the Gulf of Saint Lawrence have resulted in high mortality of NARW.

Learning about the relationship between NARW and their prey has been difficult due to the challenges associated with studying each species at suitable space and time scales \citep{mayoSurfaceForagingBehaviour1990,plourdeNorthAtlanticRight2019,meyer-gutbrodRedefiningNorthAtlantic2023}.  
In this regard, Cape Cod Bay, MA (henceforth CCB) is a critical winter and spring foraging habitat for NARW where both NARW and their prey species have been studied for decades \citep{mayoDistributionDemographyBehavior2018,clarkVisualAcousticSurveys2010a,hudakNorthAtlanticRight2023}. However, to date, challenges related to the multiple data sources and a mix of collection strategies for both NARW and their prey have limited the modeling efforts to learn about their distribution and abundance. 
Most efforts to associate NARW and their prey are at the level of the individual swimming \citep{mayoSurfaceForagingBehaviour1990} or diving whale \citep{baumgartnerSummertimeForagingEcology2003}. At larger spatial scales, research has related NARW abundance \citep{pendletonWeeklyPredictionsNorth2012a,plourdeNorthAtlanticRight2019} or calving rate \citep{meyer-gutbrodClimateassociatedChangesPrey2015} with population-level indices of prey.
The contribution of this work is to extend these efforts by building a novel joint distribution model for data collected over CCB for six days to capture this relationship and obtain more informed estimates of abundance and distribution of both NARW and their prey.

The work presented here draws on two previous efforts. One effort \citep{castillo-mateoSpacetimeMultilevelModeling2023} developed a data fusion of two types of zooplankton data obtained from differing data collection and measurement strategies, oblique net tows and surface net tows.
Zooplankton measurements made with oblique net tows are obtained by towing up through the water column, while surface net tows provide measurements of zooplankton at the surface. These methods are expected to provide insights into the actual amount of zooplankton, albeit in different ways due to the difference in towing approaches.
A second effort \citep{schliep2023assessing} focused on modeling NARW abundance. This work fused two disparate data sources, aerial distance sampling data \citep{mayoDistributionDemographyBehavior2018,Ganley2019} and acoustic monitoring data \citep{clarkVisualAcousticSurveys2010a}, both of which were also collected over CCB. This fusion utilized spatial point processes where a latent spatial point process was assumed to be generating the true point pattern (i.e., true number and location) of all NARWs in CCB. The aerial distance sampling data arise as a thinned realization of the true point pattern given the detection probabilities associated with distance sampling. For the acoustic monitoring data, points are not observed---the observed data are calls received at monitors and detection probabilities are associated with receiving calls \citep{palmerAccountingLombardEffect2022}. Further details pertaining to the zooplankton and NARW datasets are discussed with exploratory data analysis in the following section.  

Our work focuses on the joint modeling of zooplankton and NARW across CCB using a Bayesian hierarchical joint distribution model. At the data level, we combine versions of the data fusion models of \cite{castillo-mateoSpacetimeMultilevelModeling2023} and \cite{schliep2023assessing} for zooplankton and NARWs, respectively. That is, our model fuses the oblique and surface tow zooplankton data with the aerial distance sampling and passive acoustic monitoring NARW data.
In the Bayesian framework, the joint posterior distribution of the two latent processes is informed by all four datasets. The novelty of our approach is in the joint modeling at the process level of the two latent processes, one capturing zooplankton abundance, the other capturing the spatial point process intensity of NARW. 
Joint species distribution models (JSDMs) \citep[cf.][]{Tikhonov2017,isaacDataIntegrationLargeScale2020} have become a common modeling approach for multivariate species distribution data. The \emph{site level} joint process is typically specified using a latent multivariate normal distribution enabling, e.g., mean abundance or probability of presence under suitable link functions. 
Here, our joint model captures the dependence between the mean abundance of zooplankton, modeled as a geostatistical process, and the intensity of the point pattern for NARWs, formulated as a spatial point process. The model is specified in a conditional times marginal form. 
This type of joint distribution modeling hasn't been done in the literature to date but is necessitated by the different data collection for the two species. 

Specifically, our joint distribution for the latent zooplankton abundance and true intensity of the spatial point process is decomposed into the product of the marginal distribution of zooplankton and the conditional distribution of the NARW intensity given zooplankton. This choice of conditioning is two-fold. 
First, it is hard to imagine a suitable or interpretable bivariate distribution that could capture the dependence between the two disparate latent processes. Second, the conditioning enables the prediction of whale distribution and abundance given zooplankton over the study region. 
Using a comprehensive simulation study, we investigate the performance of the joint data fusion model under various sampling regimes. In particular, we study the benefits of the joint model as opposed to marginal data fusion models to estimate either the spatial distribution of zooplankton or NARW. Then, we apply our model to zooplankton and NARW data collected over CCB for six days to obtain joint estimates of their distribution and abundance and quantify the relationship between the two species. 

The remainder of the paper is as follows. Section~\ref{sec:data} provides the details on the four data sources, along with associated exploratory summaries. In Section~\ref{sec:model}, we introduce the novel joint fusion modeling of two geostatistical data sources and two point pattern data sources and provide details of inference procedure. 
Section~\ref{sec:sim} presents simulation experiments that illustrate the advantages of the joint model across various sampling scenarios and offer insights for optimizing future sampling designs. In Section~\ref{sec:real} we apply our joint fusion model to examine data collected in CCB and provide inference regarding whale and zooplankton abundance estimation. Section~\ref{sec:discussion} concludes with a discussion of the limitations of our work along with future data collection strategies that could lead to improved future inference and prediction.

\section{The data sources and exploratory analysis}
\label{sec:data}

\begin{figure}[tb]
    \centering
    \includegraphics[width = 0.9\textwidth]{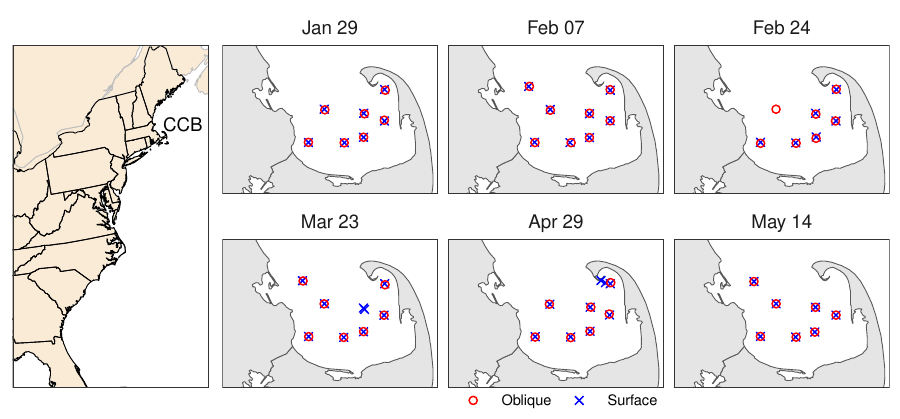}
    \caption{The location of Cape Cod Bay, MA and locations of zooplankton collection sites for each day over 6 days in 2011.}
    \label{fig:real:Zsite}
\end{figure}

In addition to being one of the most endangered large whale species on the planet, North Atlantic right whales are also one of the most well studied \citep{krausUrbanWhaleNorth2007}. Scientists have been studying NARW and their prey in Cape Cod Bay, MA, a small habitat region that NARW visit in the late winter and spring, for over 30 years \citep{mayoSurfaceForagingBehaviour1990,hudakNorthAtlanticRight2023}. Since the early 1980's the Center for Coastal Studies (CCS) has been conducting extensive net-based sampling of NARW prey throughout CCB \citep{hudakNorthAtlanticRight2023}; they have also conducted a combination of vessel- and plane-based visual surveys for NARW \citep{mayoDistributionDemographyBehavior2018,Ganley2019}. In addition, from 2007--2018, a passive acoustic array was deployed in conjunction with peak whale abundance in the Bay to capture NARW vocalizations \citep{clarkVisualAcousticSurveys2010a}. Below we detail each of these three data sources. Further details regarding the data sources and covariates, including sea surface temperature and bathymetry, used in our model are provided in the supplemental Section S1. 

\subsection{Zooplankton data}
\label{sec:data:zoop}

Starting in 1981, CCS developed a \textit{in situ} sampling program of myriad zooplankton taxa in order to better understand the prey upon which NARW feed \citep{hudakNorthAtlanticRight2023}. This program had two components: (a) a series of fixed location stations which were placed roughly evenly across CCB, designed to capture the background level of prey available over space and time and (b) a program of random sampling locations typically associated with NARW presence that was designed to better quantify the specific foraging thresholds needed to support energy gain in NARW. At each of these stations two different type of net-based sampling occurred. One was an oblique tow, whereby a sampling net (333 micron mesh) is lowered to a depth of 19m in the water column, the survey vessel gets underway, and the net is raised through the water column at an oblique angle. The second is a surface tow, whereby a 333 micron mesh net is placed just under the surface of the ocean, and is towed through a complete circle. In either type of sample, zooplankton which are caught and preserved in formalin are then taken to the lab for species identification and enumeration \citep{hudakNorthAtlanticRight2023}. The data we use here are location- and time-stamped, and are reported as total zooplankton organisms per $m^3$. Sampling is weather dependent, with the goal of sampling approximately every two weeks throughout the NARW season. We extracted zooplankton observations from six days in 2011, during which sampling was conducted across most of CCB. Figure~\ref{fig:real:Zsite} shows the locations where oblique and surface tow samplings were performed.

\subsection{Whale data}
\label{sec:data:whale}

\begin{figure}[tb]
    \centering
    \includegraphics[width = 0.9\textwidth]{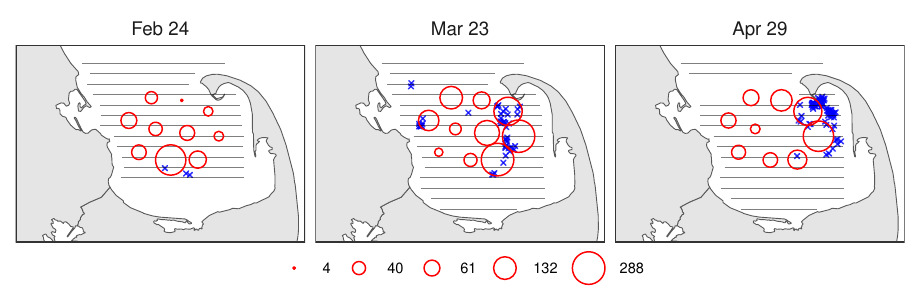}
    \caption{Observed whale locations ({\color{blue} $\times$}) by distance sampling for each day. The horizontal lines are tracklines flown. There were 3, 46, and 72 animals observed for Feb 24, Mar 23, and Apr 29, respectively. The size of {\color{red} $\circ$} represents the number of calls heard on each MARU.}
    \label{fig:real:Wdata}
\end{figure}

Following an initial pilot study in 1997, formal aerial surveys commenced in 1998 \citep{mayoDistributionDemographyBehavior2018}. The goal of these surveys was to document the distribution of NARW across CCB within the season and across years. 
Researchers also gathered information on the behavior of sighted individual whales and, through photo-identification, the identity of observed whales. 
In CCB, the surveys are comprised of a set of 15 East-West transect lines with a spacing of 2.8km (Figure~\ref{fig:real:Wdata}). Weather permitting, all 15 lines are flown in a single day; the surveys are not flown following a strict distance sampling protocol. See \citet{mayoDistributionDemographyBehavior2018} and \citet{Ganley2019} for specifics on the flight protocol, as well as information on fitted detection functions for this platform. This is discussed in more detail in Section~\ref{sec:real}. For our purposes here, we extract whale sightings from three different days in 2011, wherein most of the Bay was covered by aerial transects. Figure~\ref{fig:real:Wdata} presents locations ({\color{blue} $\times$}) where whales were seen from the plane.

In addition to the aerial surveys, the vocalizations of calling NARW have been detected with passive acoustic monitoring (PAM). Starting in 2007, and repeating annually through 2018, researchers have deployed passive acoustic arrays in CCB during the peak abundance period (approximately late February through early May) \citep{clarkVisualAcousticSurveys2010a}. The position and dimension of the array has varied slightly over the years; up through 2014 the array was wide and covered much of the bay. During the deployment, each hydrophone in a 10-element array records sound data continuously. Following retrieval, the sound data are processed, and software is used to extract NARW upcalls---a type of contact call used by all age and sex classes \citep{clarkAcousticRepertoireSouthern1982}. 
Specifically, a NARW edge-detector \citep{gillespieDetectionClassificationRight2004} enables detection and classification of NARW upcalls. Each upcall made by an individual NARW can be heard on 0 up to 10 hydrophones. 
We retain the time it was recorded, as well as the label of the hydrophone it was recorded on. In addition to upcalls, the hydrophones also record ambient noise levels, which we use in the Section~\ref{sec:model:pam} to model detection. Ambient noise in the NARW up-call frequency band (60-400 Hz) was extracted using the Filtered Noise Measurement Module from the software program PAMGuard \citep{gillespie2009pamguard} and is expressed as sound pressure level RMS in dB re: 1 \textmu Pa. 
While the hydrophones record sound continuously, for the purposes of our fusion, we only include calls that were recorded during the time when the aerial survey was flown. Figure~\ref{fig:real:Wdata} depicts the number of upcalls recorded during each unique day's flight. On 29 April 2011, we failed to record data from one hydrophone in the middle of the array.

\subsection{Exploratory data analysis}
\label{sec:eda}

\begin{figure}[tb]
    \centering
    \includegraphics[width=0.9\textwidth]{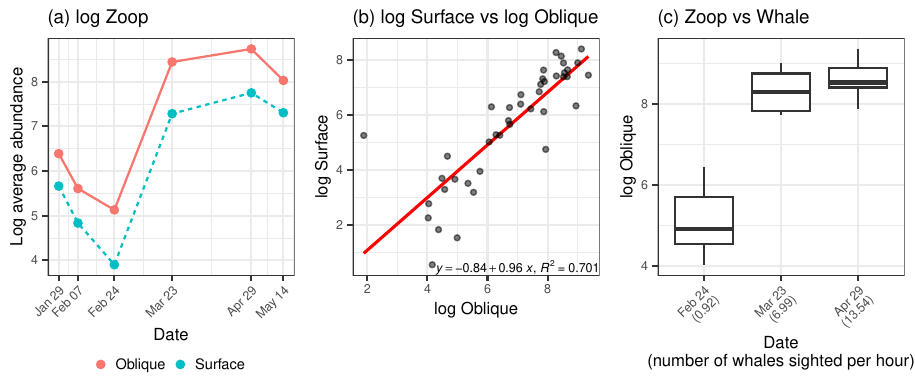}
    \caption{(a) Log average number of organisms per $m^3$ by two zooplankton measurements over 6 observation days. (b) Linear regression of log oblique on log surface. (c) Boxplot of log oblique measurements by date (the number of whales sighted by aerial survey per hour).}
    \label{fig:explor}
\end{figure}

Here, we summarize some exploratory data analysis of the zooplankton abundance per $m^3$, NARW sighting, and acoustic data to motivate building the hierarchical structure of our model. For oblique and surface tow observations, we examine kernel density estimates and normal quantile-quantile plots using the original scale and the logarithmic scale. The variance of the measurements is stabilized on the log scale, whereas they exhibit significant right skewness on the original scale. It is customary to use the log scale for zooplankton data \citep[cf.][]{plourdeNorthAtlanticRight2019,recordRapidClimateDrivenCirculation2019}

Figure~\ref{fig:explor} (a) shows changes in log average zooplankton abundance per $m^3$ for the two types of zooplankton measurements over the 6 observation days. On average, the oblique tows appear to yield larger observations compared to the surface tows. The average sea surface temperature across CCB generally increases from January through May. Zooplankton concentrations are significantly higher in spring than in winter. Figure~\ref{fig:explor} (b) presents a scatter plot between the log-transformed oblique and surface measurements. Approximate linear association appears evident suggesting linear calibration. A one-unit increase in the log oblique leads to an approximately one-unit increase in the log surface. On average, the log surface is shifted by approximately 0.8 unit below the log oblique. The log oblique explains 70\% of the variability of the log surface. Figure~\ref{fig:explor} (c) depicts the distribution of the log oblique measurements for each whale observation day. We obtain the number of whales detected per hour by dividing the detected abundance by the duration of the aerial survey. Though we have only three days, the relationship between NARW and their prey abundance seems positive and non-linear. 

\begin{table*}[tb]
\caption{The total number of calls received across hydrophones, the number of whales detected by aerial survey, duration of aerial survey in hours, and the number of whales sighted per hour. \label{tab:explor}}
\tabcolsep=0pt
\begin{tabular*}{\textwidth}{@{\extracolsep{\fill}}lrrrr@{\extracolsep{\fill}}}
\toprule%
Date & Calls & Whales sighted & Flight duration & Whales sighted per hour\\ 
  \midrule
Feb 24 & 4,147 &   3 & 3.27 & 0.92 \\ 
  Mar 23 & 6,040 &  46 & 6.58 & 6.99 \\ 
  Apr 29 & 1,936 &  72 & 5.32 & 13.54 \\ 
\bottomrule
\end{tabular*}
\end{table*}

Table~\ref{tab:explor} provides a summary of the distance sampling and PAM data. On February 24, aerial surveys recorded only 3 whale sightings, while hydrophones detected 4,147 upcalls. In contrast, on April 29, 72 whales were observed, but only 1,936 upcalls were detected. Since aerial surveys can only detect whales at the surface, this discrepancy may be attributed to temporal variation in the amount of time whales spend at the surface, changes in calling rates, or a combination of both factors.

\section{Model specification and inference}
\label{sec:model}

\subsection{Data fusion for zooplankton abundance}
\label{sec:model:zoop}

Let $\mathcal{T}_Z$ be a collection of days for which zooplankton measurements were collected. Let $Z_{t_i} (\bs)$ denote the latent number of zooplankton organisms per $m^3$ for day $t_i \in \mathcal{T}_Z$ and location $\bs \in \mathcal{D}$. For the logarithm of the latent zooplankton abundance per $m^3$, we assume
\begin{align}
    \log Z_{t_i}(\bs) &= \alpha_{0, t_i} + \alpha_{\text{temp}}\text{temp}_{t_i}(\bs) + \eta_{t_i}(\bs), \label{eq:zoop}
\end{align}
where $\alpha_{0, t_i}$'s are daily intercepts and assumed to be independent and identically distributed (i.i.d.) as a normal distribution with mean $\Tilde{\alpha}_0$ and variance $\tau^2$. The $\text{temp}_{t_i}(\bs)$ denotes the sea surface temperature for day $t_i$ and location $\bs$. The $\eta_{t_i}(\bs)$'s are daily mean 0 GPs with an exponential covariance function having variance parameter $\kappa_{\eta}$ and range parameter $\phi_{\eta}$. The daily spatial random effects are needed since the spatial distribution of zooplankton is known to substantially differ from day to day \citep{mayoSurfaceForagingBehaviour1990} and the temperature variable may not be able to adequately account for the variation. 

To inform about the latent zooplankton abundance $Z_{t_i} (\bs)$, we consider a fusion of the two data sources, oblique and surface net tows \citep{castillo-mateoSpacetimeMultilevelModeling2023}. Oblique tows are expected to have higher zooplankton abundance per $m^3$ than surface tows due to their data collection mechanism. We assume that the oblique tow captures the latent zooplankton abundance up to measurement error, whereas the surface tow requires calibration relative to the latent abundance. Let $Y^{\text{obl}}_{t_i}(\bs_j)$ denote the oblique tow observation for day $t_i \in \mathcal{T}_Z$ and location $\bs_j \in \mathcal{S}^{\text{obl}}_{t_i}$ where $\mathcal{S}^{\text{obl}}_{t_i}$ is a set of oblique tow locations for day $t_i$. For the logarithm of the oblique tow variable, we assume
\begin{align*}
    \log Y^{\text{obl}}_{t_i}(\bs_j) &= \log Z_{t_i}(\bs_j) + \epsilon^{\text{obl}}_{t_i}(\bs_j),
\end{align*}
where $\epsilon^{\text{obl}}_{t_i}(\bs_j)$'s are i.i.d. normal measurement errors with mean 0 and variance $\sigma^2_{\text{obl}}$, i.e., $\sigma^2_{\text{obl}}$ denotes the uncertainty associated with the oblique tow measurement process. The likelihood function for the oblique tows is given by
\begin{align}
    &L \left( Z_{t_i}(\bs), t_i \in \mathcal{T}_Z, \bs \in \mathcal{D} \mid \{ Y^{\text{obl}}_{t_i}(\bs_j),  t_i \in \mathcal{T}_Z, \bs_j \in \mathcal{S}^{\text{obl}}_{t_i} \} \right) \nonumber \\
    & \propto \exp\left\{ - \frac{1}{2 \sigma^2_{\text{obl}}} \sum_{t_i \in \mathcal{T}_Z} \sum_{\bs_j \in \mathcal{S}^{\text{obl}}_{t_i}} \left( \log Y^{\text{obl}}_{t_i}(\bs_j) - \log  Z_{t_i}(\bs) \right)^2 \right\}. \label{eq:lik_obl}
\end{align}

Let $Y^{\text{sur}}_{t_i}(\bs_m)$ denote the surface tow observation for day $t_i \in \mathcal{T}_Z$ and location $\bs_m \in \mathcal{S}^{\text{sur}}_{t_i}$ where $\mathcal{S}^{\text{sur}}_{t_i}$ is a set of surface tow locations for day $t_i$. We model the logarithm of the surface tow variable as 
\begin{align*}
    \log Y^{\text{sur}}_{t_i}(\bs_m) &= \lambda^{\text{sur}}_0 + \lambda^{\text{sur}}_1 \log Z_{t_i}(\bs_m) + \epsilon^{\text{sur}}_{t_i}(\bs_m),
\end{align*}
where $\lambda^{\text{sur}}_0$ and $\lambda^{\text{sur}}_1$ are calibration parameters (adopting a linear form as suggested by the EDA of the previous section).  The $\epsilon^{\text{sur}}_{t_i}(\bs_m) \stackrel{\text{i.i.d.}}{\sim} \text{N}(0, \sigma^2_{\text{sur}})$ are surface tow measurement errors with $\sigma^2_{\text{sur}}$ denoting the uncertainty associated with the surface tow measurement process. The likelihood function for the surface tow is given by
\begin{align}
    &L \left( Z_{t_i}(\bs), t_i \in \mathcal{T}_Z, \bs \in \mathcal{D} \mid \{ Y^{\text{sur}}_{t_i}(\bs_m),  t_i \in \mathcal{T}_Z, \bs_m \in \mathcal{S}^{\text{sur}}_{t_i} \} \right) \nonumber\\
    & \propto \exp\left\{ - \frac{1}{2 \sigma^2_{\text{sur}}} \sum_{t_i \in \mathcal{T}_Z} \sum_{\bs_m \in \mathcal{S}^{\text{sur}}_{t_i}} \left( \log Y^{\text{sur}}_{t_i}(\bs_m) - \lambda^{\text{sur}}_0 - \lambda^{\text{sur}}_1 \log  Z_{t_i}(\bs) \right)^2 \right\}. \label{eq:lik_sur}
\end{align}

\subsection{Data fusion for whale abundance}
\label{sec:model:whale}

Let $\mathcal{T}_W$ be a collection of days for which whale and zooplankton data were both collected with $\mathcal{T}_W$ being a subset of $\mathcal{T}_Z$. Let $\lambda^{\text{true}}_{t_{i}}(\bs)$ denote the latent true intensity of whale abundance for day $t_i \in \mathcal{T}_W$ and location $\bs \in \mathcal{D}$. For the latent true intensity, we consider the following two models.
\begin{align}
    \lambda^{\text{true}}_{t_{i}}(\bs) &= \exp \left\{ \beta_{0,t_i} + \beta_{\text{bath}}\text{bath}(\bs) + \beta_{\text{zoop}} \left( \log Z_{t_{i}}(\bs) - \overline{\log Z} \right) + \psi(\bs) \right\} \label{eq:whale1}\\
    \lambda^{\text{true}}_{t_{i}}(\bs) &= \exp \left\{ \beta_{0,t_i} + \beta_{\text{bath}}\text{bath}(\bs) + \beta_{\text{zoop}} \left( \log Z_{t_{i}}(\bs) - \overline{\log Z} \right) + \psi_{t_i}(\bs) \right\} \label{eq:whale2}
\end{align}
The $\beta_{0,t_i}$ are daily intercepts, and $\text{bath}(\bs)$ is the bathymetry at location $\bs$. The $\log Z_{t_{i}}^{\text{true}}(\bs)$ is the latent true zooplankton log abundance per $m^3$ learned from the fusion of the oblique and surface tow observations. The zooplankton covariate is centered to have mean 0 by subtracting overall average zooplankton abundance $\overline{\log Z}$ across space and time to reduce cross-correlation between the intercepts and zooplankton covariate. The model in \eqref{eq:whale1} has a Gaussian process (GP) shared across days, while the model in \eqref{eq:whale2} assumes day-specific GPs. For the GPs, we assume mean 0 and an exponential covariance function having variance parameter $\kappa_{\psi}$ and range parameter $\phi_{\psi}$. We label the joint model of \eqref{eq:zoop} and \eqref{eq:whale1} as Model (i). The joint model consisting of \eqref{eq:zoop} and \eqref{eq:whale2} is referred to as Model (ii).

Let $\mathcal{S}_{t_i}$ denote a latent point pattern realization generated from the latent true whale intensity surface $\lambda^{\text{true}}_{t_{i}}(\bs)$, i.e., $\mathcal{S}_{t_i}$ is a set of true whale locations distributed across $\mathcal{D}$ for day $t_i$. To inform about the latent true whale intensity surfaces $\lambda^{\text{true}}_{t_{i}}(\bs)$'s, we consider the fusion of two data sources, aerial distance sampling and passive acoustic monitoring \citep{schliep2023assessing}. We provide details in the following two subsections.

\subsubsection{Aerial distance sampling}
\label{sec:model:aerial}

Aerial distance sampling only provides a portion of the full point pattern $\mathcal{S}_{t_i}$. This degradation arises because the observers conducting distance sampling can only observe whales at or near the surface; they may not be able to detect individuals due to poor visibility, and the plane may not be able to survey the entire study region due to inclement weather conditions. Consider $L$ distinct aerial transects. Let $\mathcal{S}^{\text{dist}}_{t_i,\ell}$ denote a point pattern realization observed by aerial distance sampling for day $t_i \in \mathcal{T}_W$ and transect $\ell = 1,\dots,L$. We denote the degradation mechanism associated with the realization by $p^{\text{dist}}_{t_i,\ell}(\bs)$. We assume the resulting intensity $\lambda^{\text{dist}}_{t_{i}}(\bs) = p^{\text{dist}}_{t_i,\ell}(\bs) \lambda^{\text{true}}_{t_{i}}(\bs)$ has generated the observed degraded realization $\mathcal{S}^{\text{dist}}_{t_i,\ell}$ \citep{Chakraborty2011}. 

We define the degradation mechanism following \citet{Ganley2019}. Individuals on the surface are detected with probability of 1 if they are located within 0.75km of the transect. The detection probability decays exponentially when the distance exceeds 0.75km. It is assumed that $p^{\text{dist}}_{t_i,\ell}(\bs) = \pi_{t_i} f(d(\bs, \ell))$ where $\pi_{t_i}$ is the probability of a whale being on the surface for day $t_i$, $d(\bs, \ell)$ is the distance between location $\bs$ and transect $\ell$ in km, and  $f$ is a distance-based detection function given by
\begin{align*}
    f(d) = \left\{ \begin{array}{ll}
        1 & \mbox{if $d \leq 0.75$ km}  \\
        e^{-(d-0.75)^2} & \mbox{if $d > 0.75$ km}.
    \end{array} \right.
\end{align*}
Henceforth the $p^{\text{dist}}_{t_i,\ell}(\bs)$ is called a distance sampling detection probability.

We assume that the observed realizations $\mathcal{S}^{\text{dist}}_{t_i,\ell}$'s are independent across days and transects \citep{Johnson2010,Yuan2017}. The likelihood function for the distance sampling is given by
\begin{align}
    &L\left( \lambda^{\text{true}}_{t_{i}}(\bs), t_i \in \mathcal{T}_W, \bs \in \mathcal{D} \mid \{ \mathcal{S}^{\text{dist}}_{t_i,\ell}, t_i \in \mathcal{T}_W, \ell = 1,\dots,L \} \right) \nonumber\\
    & \propto \exp\left\{ - \sum_{t_i \in \mathcal{T}_W} \sum_{\ell=1}^L \int_{\mathcal{D}} \lambda^{\text{dist}}_{t_{i}}(\bs) d \bs \right\} \prod_{t_i \in \mathcal{T}_W} \prod_{\ell=1}^L \prod_{\bs_j \in \mathcal{S}^{\text{dist}}_{t_i,\ell}} \lambda^{\text{dist}}_{t_{i}}(\bs_j). \label{eq:lik_dist}
\end{align}

\subsubsection{Passive acoustic monitoring}
\label{sec:model:pam}

The PAM data source we used does not provide whale locations but provides a set of times at which NARW upcalls were received at each hydrophone $k$ for $k = 1, \dots, K$. For each day $t_i \in \mathcal{T}_W$, we obtain the number $W_{t_i, k}$ of upcalls received at each hydrophone $k$ during the time window of the aerial survey conducted on that day. The PAM data also exhibit degradation due to our ability to detect upcalls; this degradation may arise because the calling whale is far from a hydrophone, the source level of its call is low, and/or the background ambient noise level is high, thereby reducing the signal to noise ratio. Let $p^{\text{PAM}}_{t_i,k}(\bs)$ denote the degradation mechanism associated with the realization $W_{t_i, k}$. We assume that $W_{t_i, k}$ is a realization from a Poisson distribution with mean $\lambda^{\text{PAM}}_{t_i, k} = c_{t_i} d_{t_i} \int_{\mathcal{D}} p^{\text{PAM}}_{t_i,k}(\bs) \lambda^{\text{true}}_{t_i}(\bs) d \bs$ where $c_{t_i}$ is the number of upcalls made per whale per hour for day $t_i$, $d_{t_i}$ is the aerial survey duration in hours, and $\int_{\mathcal{D}} p^{\text{PAM}}_{t_i,k}(\bs) \lambda^{\text{true}}_{t_i}(\bs) d \bs$ represents the expected number of whales whose calls can be detected by hydrophone $k$.

The degradation mechanism we adopt follows from \citet{palmerAccountingLombardEffect2022} and \citet{schliep2023assessing}. The $p^{\text{PAM}}_{t_i,k}(\bs)$ represents the probability of hydrophone $k$ detecting a call made at location $\bs$ for day $t_i$. Let $\text{noise}_{t_i, k}$ denote the median ambient noise level in dB re:1$\mu$Pa at hydrophone $k$ during aerial survey for day $t_i$. An upcall source level (SL) is assumed to be uniformly distributed over the interval [141 dB, 197 dB]. The upcall SL must be 26 dB above $\text{noise}_{t_i, k}$ for hydrophone $k$ to detect it. The estimated transmission loss function is given by $14.5 \log_{10}(d(\bs, k))$ where $d(\bs, k)$ is the distance in meters between location $\bs$ and hydrophone $k$. Altogether, the degradation mechanism for the PAM data source is defined as
\begin{align*}
    &p^{\text{PAM}}_{t_i,k}(\bs)\\
    &= \text{P} \left(\text{SL} - 14.5 \log_{10}(d(\bs, k)) > \text{noise}_{t_i, k} + 26 \right)\\
    & = 1 - \text{P} \left( \text{SL} \leq 14.5 \log_{10}(d(\bs, k)) + \text{noise}_{t_i, k} + 26 \right)\\
    & = \left\{ \begin{array}{ll}
        0 & \mbox{if } 14.5 \log_{10}(d(\bs, k)) + \text{noise}_{t_i, k} + 26 > 197 \\
        1 & \mbox{if } 14.5 \log_{10}(d(\bs, k)) + \text{noise}_{t_i, k} + 26 < 141 \\
        1 - \frac{ 14.5 \log_{10}(d(\bs, k)) + \text{noise}_{t_i, k} + 26 - 141 }{197 - 141} & \mbox{otherwise}.
    \end{array} \right. 
\end{align*}
We call $p^{\text{PAM}}_{t_i,k}(\bs)$ a PAM detection probability hereafter. 

We assume the observed realizations $W_{t_i,k}$ are independent across days and hydrophones. The likelihood function for the PAM has the form
\begin{align}
    &L\left( \lambda^{\text{true}}_{t_{i}}(\bs), t_i \in \mathcal{T}_W, \bs \in \mathcal{D} \mid \{ W_{t_i,k}, t_i \in \mathcal{T}_W, k = 1, \dots, K \} \right) \nonumber\\
    & \propto \exp \left\{ \sum_{t_i \in \mathcal{T}_W} \sum_{k=1}^K \lambda^{\text{PAM}}_{t_i, k} \right\} \prod_{t_i \in \mathcal{T}_W} \prod_{k=1}^K ( \lambda^{\text{PAM}}_{t_i, k} )^{W_{t_i, k}}. \label{eq:lik_pam}
\end{align}

\subsection{Bayesian inference}
\label{sec:infer}

\begin{figure}[tb] 
    \centering
    \caption{An illustrative diagram for the joint zooplankton-whale models. Solid-lined blocks denote observed data sources, whereas dashed-lined blocks represent latent variables or model parameters. The edges show the relationships between the different components.}
    \label{fig:model}
    \begin{tikzpicture}[node distance=1cm, auto, thick, >=Stealth]
    
    \tikzstyle{data} = [draw, rectangle, rounded corners, text centered, minimum height=1cm, minimum width=1cm, font=\small]
    \tikzstyle{latent} = [draw, rectangle, rounded corners, text centered, minimum height=1cm, minimum width=1cm, font=\small, dashed]
    \tikzstyle{arrow} = [thick,->,>=stealth]
    
    \node (temp) [data] {temp$_{t_i}(\bs)$};
    \node (alpha) [latent, right = 0.5cm of temp] {$\alpha_{0, t_i}, \alpha_{temp}, \eta_{t_i}(\bs)$};
    \node (z_true) [latent, below = 0.8cm of temp, xshift=1.3cm] {$Z_{t_i}(\bs)$};
    \node (yobl) [data, below = 2.6cm of temp] {$Y^{obl}_{t_i}(\bs_j)$};
    \node (ysur) [data, right = 0.6cm of yobl] {$Y^{sur}_{t_i}(\bs_m)$};
    \node (par_obl) [latent, below = 0.8cm of yobl] {$\sigma^2_{obl}$};
    \node (par_sur) [latent, below = 0.8cm of ysur] {$\lambda^{sur}_0, \lambda^{sur}_1, \sigma^2_{sur}$};
    
    \node (bath) [data, right = 1cm of alpha] {bath$(\bs)$};
    \node (beta) [latent, right = 3cm of alpha] {$\beta_{0, t_i}, \beta_{bath}, \beta_{zoop}, \psi(\bs)$ or $\psi_{t_i}(\bs)$};
    \node (lambda_true) [latent, below = 0.8cm of bath, xshift=1.8cm] {$\lambda^{true}_{t_i}(\bs)$};
    \node (dist) [latent, below = 2.6cm of bath, xshift=0.8cm] {$\lambda^{dist}_{t_i,\ell}(\bs)$};
    \node (pam) [latent, right = 0.6cm of dist] {$\lambda^{PAM}_{t_i,k}$};
    \node (pi) [latent, left=0.6cm of dist] {$\pi_{t_i}$};
    \node (c) [latent, right=0.7cm of pam, yshift=0.7cm] {$c_{t_i}$};
    \node (noise) [data, right=0.7cm of pam, yshift=-0.7cm] {noise$_{t_i,k}$};
    \node (s_dist) [data, below=0.8cm of dist] {$S^{dist}_{t_i, \ell}$};
    \node (w) [data, below=0.8cm of pam] {$W_{t_i, k}$};
    
    \draw [arrow] (temp) -- (z_true);
    \draw [arrow] (alpha) -- (z_true);
    \draw [arrow] (z_true) -- (yobl);
    \draw [arrow] (z_true) -- (ysur);
    \draw [arrow] (par_obl) -- (yobl);
    \draw [arrow] (par_sur) -- (ysur);
    
    \draw [arrow] (bath) -- (lambda_true);
    \draw [arrow] (lambda_true) -- (dist);
    \draw [arrow] (lambda_true) -- (pam);
    \draw [arrow] (dist) -- (s_dist);
    \draw [arrow] (pam) -- (w);
    \draw [arrow] (beta) -- (lambda_true);
    \draw [arrow] (pi) -- (dist);
    \draw [arrow] (c) -- (pam);
    \draw [arrow] (noise) -- (pam);

    \draw [arrow] (z_true) -- (lambda_true);
    
    \end{tikzpicture}

\end{figure}
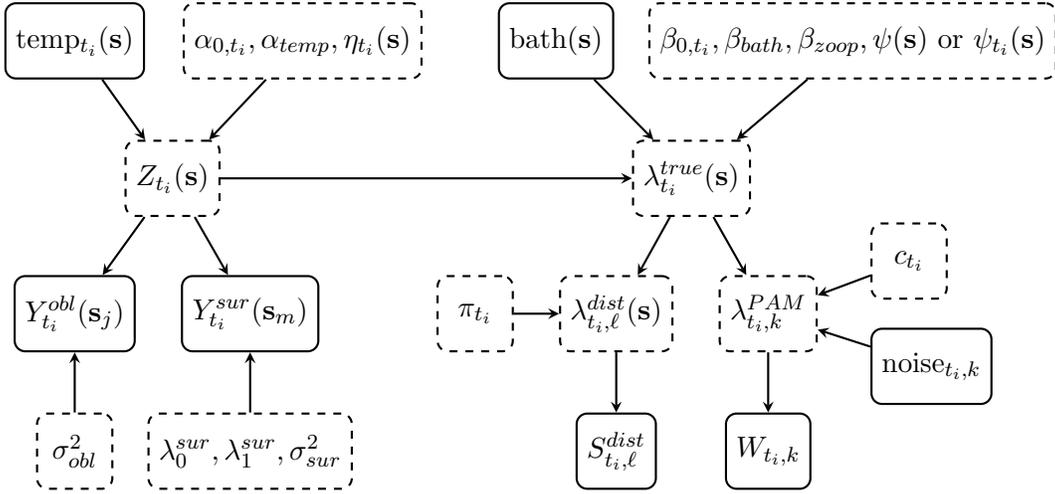

A schematic representation of our proposed joint zooplankton-whale models is provided in Figure~\ref{fig:model}. The solid-line blocks correspond to observed data, while dashed-line blocks represent latent variables or parameters that are inferred through the models. This demonstrates the flow of information between the different components of the model. Environmental covariates, such as sea surface temperature and bathymetry, influence the latent zooplankton and whale abundance. Observed data on zooplankton and whale abundance ($Y^{\text{obl}}_{t_i} (\bs_j)$, $Y^{\text{sur}}_{t_i} (\bs_m)$, $\mathcal{S}^{\text{dist}}_{t_i}$ and $W_{t_i, k}$) are used to estimate the latent states of these populations. 
Additionally, the model incorporates other parameters and covariates, such as ambient noise levels ($\text{noise}_{t_i,k}$), surface time probabilities ($\pi_{t_i}$), and call rates ($c_{t_i}$), which affect the distribution and movement of whales and zooplankton across space and time.
The likelihood function for the joint model is the product of \eqref{eq:lik_obl}, \eqref{eq:lik_sur}, \eqref{eq:lik_dist}, and \eqref{eq:lik_pam}.



We assume independent multivariate normal priors with mean of \textbf{0} and covariance matrix of 100\textbf{I} for $\balpha^{\top} = (\alpha_{0, t_1}, \dots, \alpha_{0, t_{N(\mathcal{T}_Z)}}, \alpha_{\text{temp}})$ and $\bbeta^{\top} = (\beta_{0, t_1}, \dots, \beta_{0, t_{N(\mathcal{T}_W)}}, \beta_{\text{bath}}, \beta_{\text{zoop}})$ where $N(\mathcal{T})$ denotes the number of elements for a set $\mathcal{T}$. We assign independent normal priors with mean of 0 and variance of 100 to $\Tilde{\alpha}_{0}$, $\mu_0$, $\lambda^{\text{sur}}_0$, and $\lambda^{\text{sur}}_1$. The variance parameters, $\sigma^2_{\text{sur}}$, $\kappa_{\eta}$, $\kappa_{\delta}$, and $\kappa_{\psi}$, are assigned inverse gamma priors with shape and rate parameters equal to 2. For computational efficiency, we fix the spatial decay parameters, $\phi_{\eta}$ and $\phi_{\psi}$, to an effective range equal to $\frac{1}{4}$ max distance, which is approximately equal to 16 km. We assign independent gamma priors to the call rates $c_{t_i}$, with a shape parameter of 9/4 and a scale parameter of 10/7.5, resulting in a mean of 3 and a variance of 4. This choice is informed by auxiliary data from high-resolution tag deployments conducted by S. Parks and D. Wiley (pers. comm) in April 2009 and 2010. The data consist of observed call rates per hour for three whales during a sampling event, with values of 0.6, 3.24, and 3.54, respectively.

The proposed joint models can account for daily-varying spatial distributions of zooplankton and NARWs and provide inference on the relationship between NARW abundance and prey availability. However the datasets we consider may not be large enough or have adequate spatial coverage to train the proposed complicated model. For instance, the zooplankton observations were collected at at most 9 locations a day, which is very few to learn about the spatial Gaussian processes. Thus we assume that some parameters are known and fix them to estimates obtained by external auxiliary data or extensively referenced literature. For the zooplankton model, we fix the variance $\tau^2$ of daily intercepts at 0.04 to ensure that $\log Z_{t_i}(\bs)$ ranges from 0 to 10, as observed in the oblique observations.  The monthly surface time probabilities are fixed as $\pi_{Feb} = 0.34$, $\pi_{Mar} = 0.31$ and $\pi_{Apr} = 0.55$ following \citet{Ganley2019}. 
We fix the oblique tow measurement error $\sigma^2_{\text{obl}}$ to 1, based on external auxiliary data provided by C. Hudak (pers. comm). This data set includes 15 pairs of oblique tow samples and measurements presumed to reflect the latent zooplankton abundance. The sample variance of their log ratios is used for $\sigma^2_{\text{obl}}$.


We use Markov chain Monte Carlo (MCMC) for model fitting and generate posterior samples of model parameters. To evaluate the stochastic integrals, we use numerical integration where the area of CCB is discretized into approximately 2,000 grid cells of dimension of 1km $\times$ 1km. For simplification, let $D_W$ and $\btheta_W$ be the data and parameters for the whale model, respectively. Let $D_Z$ and $\btheta_Z$ be the data and parameters for the zooplankton model, respectively. The full data model can be expressed as $[D_W \mid \btheta_W, Z] [D_Z \mid \btheta_Z, Z]$. The posterior distribution is given by $[\btheta_W, \btheta_Z, Z \mid D_W, D_Z]$. At each MCMC step, we update the unknowns as follows.
\begin{align*}
    &\text{(i) } [\btheta_W \mid D_W, D_Z, Z, \btheta_Z] \propto [D_W \mid Z, \btheta_W] [\btheta_W]\\
    &\text{(ii) } [Z \mid D_W, D_Z, \btheta_W, \btheta_Z] \propto [D_W \mid Z, \btheta_W] [D_Z \mid Z, \btheta_Z] [Z \mid \btheta_Z]\\
    &\text{(iii) } [\btheta_Z \mid D_W, D_Z, Z,\btheta_W] \propto [D_Z \mid Z, \btheta_Z] [Z \mid \btheta_Z] [\btheta_Z]
\end{align*}
However, updating the latent latent zooplankton abundance $Z_{t_i}(\bs)$ at each step can make the algorithm computationally expensive and result in slow mixing. 
We marginalize out $Z_{t_i}(\bs)$ from the posterior distribution as $[\btheta_W, \btheta_Z \mid D_W, D_Z] = \int [\btheta_W, \btheta_Z, Z \mid D_W, D_Z] d Z$ and update the unknowns as follows.
\begin{align*}
    &\text{(i) } [\btheta_W \mid D_W, D_Z, \btheta_Z] \propto [D_W \mid \btheta_W, \btheta_Z] [\btheta_W] \nonumber \\
    &\text{(ii) } [\btheta_Z \mid D_W, D_Z,\btheta_W] \propto [D_W \mid \btheta_W, \btheta_Z] [D_Z \mid \btheta_Z] [\btheta_Z]
\end{align*}

\section{Simulation experiments}
\label{sec:sim}

An aim of these simulation experiments is to examine abundance recovery for both zooplankton and NARW as well as to demonstrate what benefits exist and when by jointly modeling NARW abundance and prey concentration. We first elaborate a process of simulating the four different data sources, i.e., oblique tow, surface tow, distance sampling, and acoustic detection datasets. These simulations are designed to provide latent abundance and sampled observations analogous to our real-world scenario, informed by the sampling protocols associated with each data source. We examine how well the two different models recover latent abundance of zooplankton and whales. We vary the features of data sources in order to learn how the performance of our proposed models varies across a range of scenarios. 

\subsection{Simulation of data sources}
\label{sec:sim:data}

We consider six days of zooplankton sampling and three days of whale data collection, consistent with the real CCB data sources. To simulate \textit{observed} zooplankton and whale data, we start with spatially explicit \textit{latent} average zooplankton abundance $Z_{t_i}(\bs)$ and \textit{latent} whale intensity $\lambda^{\text{true}}_{t_i}(\bs)$ for $\bs \in \mathcal{D}$ where $\mathcal{D}$ is the area of CCB. The spatial window $\mathcal{D}$ is discretized using a 1km $\times$ 1km resolution grid. We consider actual covariate variables (i.e., sea surface temperature and bathymetry) collected across CCB for data generation. For all the covariates and GPs, we obtain associated values at the collection of grid cell centroids. The true values of the parameters for the zooplankton models are $\Tilde{\alpha}_0 = 5.5$, $\tau^2 = 0.04$, $\alpha_{\text{temp}} = 0.2$, $\kappa_{\eta} = 1$, $\sigma^2_{\text{sur}} = 0.5$, $\lambda^{\text{sur}}_0 = -0.7$, and $\lambda^{\text{sur}}_1 = 1$. The $\alpha_{0, t_1}, \dots, \alpha_{0, t_6}$ are independently sampled from N$(\Tilde{\alpha}_0, \tau^2)$. For the whale components, we set $\bbeta^\top = (\beta_{0,t_3}, \beta_{0,t_4}, \beta_{0,t_5}, \beta_{\text{bath}}, \beta_{\text{zoop}}) = (-4, -4, -4, -0.26, 1.5)$, $\kappa_{\psi} = 0.2$, and $c_{t_3} = c_{t_4} = c_{t_5} = 2.46$.

For simplification, we describe the data simulation process for a given day $t_i$. Given the latent average zooplankton abundance surface, for each collection site $\bs_j$, we find a grid cell having $\bs_j$ and use the latent average abundance associated with the grid cell to generate oblique and surface tow observations. 

Given the true whale intensity surface, we simulate a point pattern realization of true whale locations \citep{gelfand2018bayesian}. For each transect $\ell$, the collection $\mathcal{S}^{\text{dist}}_{t_i,\ell}$ of distance sampling observations is simulated based on the detection function $p^{\text{dist}}_{t_i, \ell}(\bs)$ and true whale locations. For each whale location $\bs_i$, we evaluate a detection probability $p^{\text{dist}}_{t_i, \ell}(\bs_i)$ using the distance between $\bs_i$ and the nearest point along the transect $\ell$ and generate a binary random variable from a Bernoulli distribution having the detection probability as its mean. The binary variable is 1 if the whale is detected by the transect $\ell$, and it is 0 otherwise. The $\mathcal{S}^{\text{dist}}_{t_i,\ell}$ comprises the locations of whales detected by the distance sampling associated with transect $\ell$. Combining $\mathcal{S}^{\text{dist}}_{t_i,\ell}$ across $\ell=1,2,...L$ yields total observed whale locations from the aerial survey.

To simulate PAM data, we first generate the number of calls each whale makes independently from a Poisson distribution with mean equal to $c_{t_i} d_{t_i}$ which is the average number of calls per whale during distance sampling. The total number $W_{t_i, k}$ of calls heard on each hydrophone $k$ is simulated based on the PAM detection probability $p^{\text{PAM}}_{t_i, k}(\bs)$ and true whale locations. For every call originated from each whale location $\bs_i$, we compute a detection probability $p^{\text{PAM}}_{t_i, k}(\bs_i)$ using the noise level at hydrophone $k$ and the distance between $\bs_i$ and hydrophone $k$. Then, we independently generate a binary random variable from a Bernoulli distribution with the mean equal to the detection probability. The binary variable indicates whether or not hydrophone $k$ detects the call. Summing over these binary variables provides the total number $W_{t_i, k}$ of calls detected at hydrophone $k$.

\subsection{Results}
\label{sec:sim:res}

\begin{figure}[tb]
    \centering
    \includegraphics[width = 0.9\textwidth]{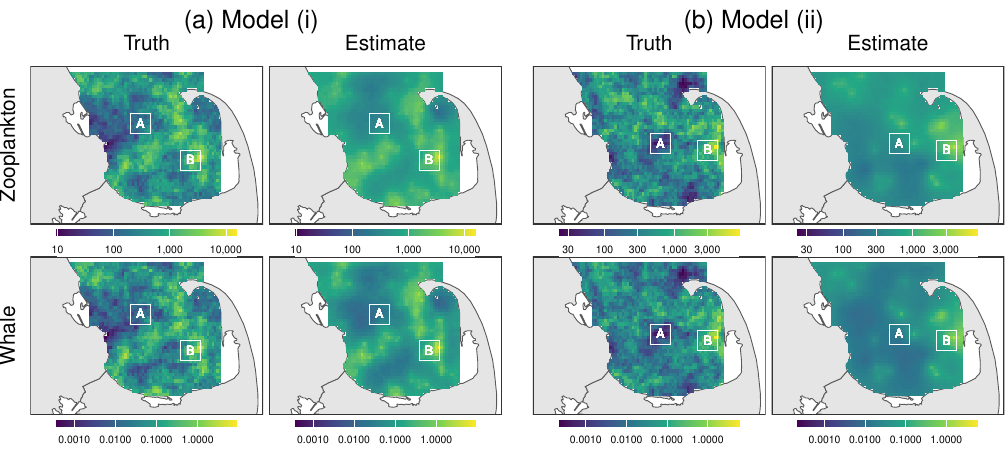}
    \caption{(a) Latent and posterior mean estimated surfaces of zooplankton abundance per $m^3$ (top) and whale intensity (bottom) for Model (i) fitted to data simulated from the model. (b) Those for Model (ii) fitted to data simulated from the model.}
    \label{fig:sim:abunM1M2}
\end{figure}

We consider real zooplankton collection locations (Figure~\ref{fig:real:Zsite}) and real flight paths and arrays of active hydrophones (Figure~\ref{fig:real:Wdata}) to simulate data from each of Models (i) and (ii). Given the likelihood functions in \eqref{eq:lik_obl}, \eqref{eq:lik_sur}, \eqref{eq:lik_dist}, and \eqref{eq:lik_pam}, each of Models (i) and (ii) was fitted to data generated from the model (i.e., the generating model is identical to the fitting model). For each model, we ran three MCMC chains with disparate starting values. Each of the chains was run for 150,000 iterations until convergence, and we use the last 50,000 iterations for inference. 
Using Intel Xeon Gold 6252 CPUs on the Duke Compute Cluster, the model (i) took approximately 1.7 days and the model (ii) took approximately 2.5 days. 
In this section, we only display results associated with day $t_5$ for simplicity. Similar results are observed for other days and can be found in the supplemental Section~S2. 

\begin{table*}[tb]
\caption{Latent values, posterior median estimates, and 95\% HPD intervals of average zooplankton abundance per $m^3$ (top) and whale abundance (bottom) for the entire study region ($\mathcal{D}$) and two subregions (A and B) shown in Figure~\ref{fig:sim:abunM1M2} for Models (i) and (ii) fitted to their respective simulated data. \label{tab:sim:abun12}}
\tabcolsep=0pt
\begin{tabular*}{\textwidth}{@{\extracolsep{\fill}}ll rrr rrr@{\extracolsep{\fill}}}
  \toprule%
  \multirow{2}{*}{Component} & \multirow{2}{*}{Region} & \multicolumn{3}{@{}c@{}}{Model (i)} & \multicolumn{3}{c}{Model (ii)}\\
  \cmidrule{3-5} \cmidrule{6-8}%
   & & Truth & Median & 95\% HPD & Truth & Median & 95\% HPD \\ 
  \midrule
  Zooplankton  & $\mathcal{D}$ & 682 & 485 & (138, 1346) & 546 & 542 & (199, 1118) \\ 
   & A & 248 & 210 & (41, 645) & 144 & 388 & (82, 977) \\ 
   & B & 1847 & 943 & (156, 3523) & 1501 & 1205 & (309, 3253) \\
  \midrule
  Whale & $\mathcal{D}$ & 241 & 264 & (212, 323) & 94 & 78 & (47, 117) \\ 
   & A & 0 & 1 & (0, 3) & 0 & 1 & (0, 2) \\ 
   & B & 30 & 38 & (26, 53) & 16 & 15 & (6, 26) \\ 
  \bottomrule
\end{tabular*}
\end{table*}

Figure~\ref{fig:sim:abunM1M2} shows latent and posterior mean estimated surfaces of zooplankton abundance per $m^3$ and whale intensity for each model. Overall, the estimated surfaces exhibit trends that are comparable to the truth across the region. For each model, we compute the posterior estimates of zooplankton abundance per $m^3$ across the entire study region $\mathcal{D}$ by averaging the posterior sample of $Z(\bs)$ evaluated at each grid cell. The total whale abundance estimates are obtained by numerically integrating the posterior sample of $\lambda(\bs)$ over $\mathcal{D}$. To evaluate the local performance of the proposed models, we also estimate the abundance for subregions A and B, selected for their differences in abundance, shown in Figure~\ref{fig:sim:abunM1M2}. Table~\ref{tab:sim:abun12} presents true values, posterior median estimates, and 95\% HPD intervals for the abundance for the entire region $\mathcal{D}$ and two subregions A and B. Both models provide the posterior median estimates similar to the truth, and their 95\% HPD intervals capture the truth for every case. 
The estimated zooplankton abundance carries significant uncertainties. This is likely due to the insufficient number of observations (at most 9 locations a day) compared to the complexity of the zooplankton model, which incorporates daily GPs to account for anticipated day-to-day spatial variation.

\begin{figure}[tb]
    \centering
    \includegraphics[width = 0.75\textwidth]{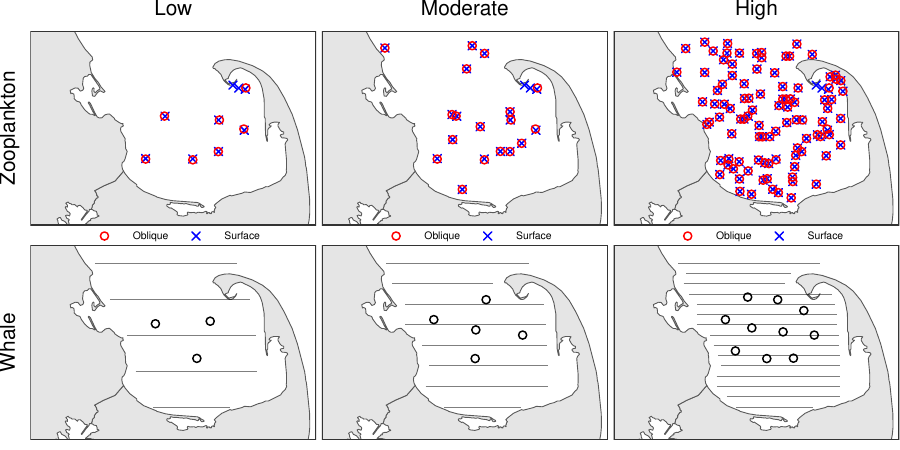}
    \caption{Low, moderate, and high sampling intensity for zooplankton data collection (top) and whale observation (bottom).}
    \label{fig:sim:samp}
\end{figure}

To further study the performance of our proposed models, we consider several different scenarios of data collection design. For the zooplankton collection, we consider low, moderate, and high sampling intensities defined by the real zooplankton sites, 20 sites per day, and 100 sites per day, respectively, which are displayed in the top of Figure~\ref{fig:sim:samp}. For the whale monitoring, we consider low, moderate, and high sampling intensities designed by 5 transects and 3 hydrophones, 8 transects and 5 hydrophones, and 15 transects and 10 hydrophones, respectively, which are depicted in the bottom of Figure~\ref{fig:sim:samp}. Note that the low zooplankton sampling intensity and high whale sampling intensity are analogous to the sampling design of the real data. 

\begin{table*}[tb]
\caption{Posterior median estimates (95\% HPD intervals) of $\beta_{\text{zoop}}$ and average zooplankton abundance per $m^3$ across $\mathcal{D}$, RMSE of log zooplankton abundance surface, and CRPS of average zooplankton abundance across $\mathcal{D}$ under different whale sampling scenarios. The true values of $\beta_{\text{zoop}}$ and average zooplankton abundance are 1.5 and 682, respectively. \label{tab:sim:Zabun_incWhale}}
\tabcolsep=0pt
\begin{tabular*}{\textwidth}{@{\extracolsep{\fill}}llrrrr@{\extracolsep{\fill}}}
\toprule%
  Model & Whale sampling & $\beta_{\text{zoop}}$ & Zoop abundance & RMSE & CRPS \\ 
  \midrule
  Zooplankton & None & - & 1014 (485, 1805) & 0.97 & 212.0 \\ 
  \midrule
  \multirow{3}{*}{Joint} & Low & 1.54 (0.90, 2.35) & 820 (445, 1317) & 0.89 & 86.9 \\ 
  & Moderate & 1.09 (0.62, 1.57) & 795 (438, 1305) & 0.75 & 76.4 \\ 
  & High & 1.40 (1.03, 1.75) & 772 (413, 1274) & 0.73 & 64.5 \\
\bottomrule
\end{tabular*}
\end{table*}

We simulate oblique and surface tow observations under the moderate zooplankton sampling intensity. Given true whale intensity surfaces, we simulate aerial and acoustic data under each of the low, moderate, and high whale sampling intensities. We fit a zooplankton model (left side of Figure~\ref{fig:model}) to the simulated zooplankton data and compare its abundance estimation performance with that of our model (entire Figure~\ref{fig:model}) jointly fitted to the zooplankton and whale data under the different whale sampling designs. For the comparison, we compute the posterior median estimate and 95\% HPD interval for average zooplankton abundance per $m^3$ across $\mathcal{D}$. The root mean squared error (RMSE) of the log average zooplankton abundance surface is also evaluated using the discretized latent and estimated zooplankton abundance surfaces. Additionally we calculate the continuous ranked probability score (CRPS) based on the posterior samples for the average zooplankton abundance per $m^3$ across $\mathcal{D}$. For the joint model, we compute the posterior median estimates and 95\% HPD intervals of $\beta_{\text{zoop}}$, which infers the relationship between whale abundance and zooplankton concentration, for the three different sampling scenarios. These values are displayed in Table~\ref{tab:sim:Zabun_incWhale}. Across all sampling scenarios, the 95\% HPD interval successfully captures the true value of $\beta_{\text{zoop}}$. Additionally, the uncertainty in the estimates decreases as the sampling intensity increases. Jointly modeling zooplankton and whale variables produces posterior median estimates that more closely align with the true values, exhibiting reduced uncertainty, as well as lower RMSE and CRPS, in comparison to the case that incorporates only zooplankton data. Additionally, increasing the intensity of whale sampling results in estimates that are more accurate, with further reductions in RMSE and CRPS. This demonstrates that our joint modeling framework enhances the estimation of zooplankton abundance, with improved accuracy and precision as the whale sampling intensity increases.

\begin{table*}[tb]
\caption{Posterior median estimates (95\% HPD intervals) of $\beta_{\text{zoop}}$ and total whale abundance, RMSE of log whale intensity surface, and CRPS of total whale abundance. The true values of $\beta_{\text{zoop}}$ and total whale abundance are 1.5 and 241, respectively. \label{tab:sim:Wabun_incZoop}}
\tabcolsep=0pt
\begin{tabular*}{\textwidth}{@{\extracolsep{\fill}}llrrrr@{\extracolsep{\fill}}}
\toprule%
 Model & Zoop sampling & $\beta_{\text{zoop}}$ & Whale abundance  & RMSE & CRPS \\ 
  \midrule
 Whale & None & - & 215 (160, 277) & 1.25 & 15.4 \\ 
   \midrule
  \multirow{3}{*}{Joint} & Low & 1.67 (0.91, 2.54) & 214 (154, 285) & 1.18 & 16.0 \\ 
   & Moderate & 1.09 (0.62, 1.57) & 251 (187, 324) & 1.10 & 9.1 \\ 
   & High & 1.23 (0.80, 1.66) & 237 (175, 305) & 1.05 & 8.0 \\ 
\bottomrule
\end{tabular*}
\end{table*}

Based on the latent zooplankton abundance surfaces, we simulate oblique and surface tow observations for low, moderate, and high zooplankton sampling intensities. We use moderate whale sampling intensity to simulate aerial and acoustic observations. We apply a whale model (right side of Figure~\ref{fig:model}) to the simulated whale data and compare its performance in estimating whale abundance with our joint model (entire Figure~\ref{fig:model}), which is fitted to both zooplankton and whale data across various zooplankton sampling designs. 
For this comparison, we calculate the posterior median estimate and 95\% HPD interval for total whale abundance. We also assess the RMSE of the log whale intensity using the discretized true and estimated whale intensity surfaces. Additionally, the CRPS is computed from the posterior samples for total whale abundance. For the joint model, 
we compute the posterior median estimates and 95\% HPD intervals of $\beta_{\text{zoop}}$ across the three different sampling scenarios. The results of these metrics are presented in Table~\ref{tab:sim:Wabun_incZoop}. 
Our joint model consistently captures the true value of $\beta_{\text{zoop}}$ across all scenarios based on the 95\% HPD intervals, and the uncertainty decreases with higher sampling intensity. 
Our joint model provides more accurate estimates of whale abundance and intensity surface compared to the whale model, as evidenced by median estimates that are closer to the true values, lower uncertainty, and reduced RMSE and CRPS.
Furthermore, higher zooplankton sampling intensity results in more accurate estimates, with additional reductions in RMSE and CRPS. This highlights that our joint modeling framework is expected to improve the accuracy and precision of whale abundance estimates as zooplankton sampling intensity increases.


\section{Application to CCB data sources}
\label{sec:real}



We apply our joint models to data collected on North Atlantic right whales in Cape Cod Bay in 2011 described in Section~\ref{sec:data}. As noted briefly in Section~\ref{sec:data:whale}, the aerial surveys were not conducted in strict accordance with the standard distance sampling protocol where observers follow straight lines (i.e., transects) and record the perpendicular distances from the line to each detected individual or group of animals. Supplemental Figure S6 depicts the actual flight path of the plane during the aerial survey on each observation day. When observers spotted an individual or a group of animals, they approached and circled above the location and recorded the number of observed individuals (i.e., multiple whales can be recorded at the same location). This could result in detecting more individuals than anticipated under a strict distance sampling protocol. Therefore, in our analysis, we employed two different 
repairs. One assumes that the surveys adhered to the strict distance sampling protocol and include all recorded whale sightings. This approach provides an upper bound on estimation of whale abundance. The other assumes that only one individual was observed at each location, regardless of how many whales were actually sighted. This approach yields a lower bound on estimation of whale abundance. The results from the former approach are presented here, with most of the results from the latter approach available in the supplemental Section S3.  The model fitting was carried out using Markov chain Monte Carlo, with a total of 200,000 iterations until convergence. We retained the last 50,000 iterations for inference. 

\begin{table*}[tb]
\caption{Posterior median estimate and 95\% HPD interval for the loglikelihood function as well as CRPS for whale abundance detected by distance sampling for each day. \label{tab:real:comp}}
\tabcolsep=0pt
\begin{tabular*}{\textwidth}{@{\extracolsep{\fill}}l rr rrr@{\extracolsep{\fill}}}
\toprule%
\multirow{2}{*}{Model} & \multicolumn{2}{c}{Loglikelihood} & \multicolumn{3}{c}{CRPS}\\
\cmidrule{2-3} \cmidrule{4-6}%
  & Median & 95\% HPD & Feb 24 & Mar 23 & Apr 29 \\
  \midrule
  (i) & 10747 & (10723, 10769) & 2.19 & 1.58 & 2.03 \\ 
  (ii) & 10746 & (10720, 10772) & 2.21 & 1.58 & 2.06 \\ 
\bottomrule
\end{tabular*}
\end{table*}

For model comparison, the continuous ranked probability score (CRPS) is calculated based on the posterior samples for the number of whales detected by aerial distance sampling for each day for each model. Additionally, we compare the posterior distributions of the loglikelihood function between Models (i) and (ii). The results, presented in Table~\ref{tab:real:comp}, indicate that the posterior median estimates of the loglikelihood function are comparable between the two models. However, Model (ii) exhibits greater uncertainty, likely due to the additional variability introduced by the daily GPs associated with the whale component. Model (i) produces CRPS values that are slightly smaller than or equal to those of Model (ii). Collectively, these metrics suggest that the models provide comparable performance in fitting our dataset. We present further model inference based on Model (i) since it is simpler but performs as well as Model (ii).

\begin{table*}[tb]
\caption{Estimated posterior medians and 95\% HPD intervals for model parameters. \label{tab:real:par}}
\tabcolsep=0pt
\begin{tabular*}{\textwidth}{@{\extracolsep{\fill}}lrr lrr@{\extracolsep{\fill}}}
\toprule%
\multicolumn{3}{c}{Zooplankton} & \multicolumn{3}{c}{Whale}\\
\cmidrule{1-3} \cmidrule{4-6}%
Parameter & Median & 95\% HPD & Parameter & Median & 95\% HPD\\
\midrule
  $\alpha_{\text{temp}}$ & 0.49 & (-0.76, 1.49) & $\beta_{\text{bath}}$ & 0.10 & (-1.10, 1.26) \\ 
  $\lambda^{\text{sur}}_{0}$ & -1.85 & (-3.80, 0.06) & $\beta_{\text{zoop}}$ & 4.13 & (3.24, 5.05) \\ 
  $\lambda^{\text{sur}}_{1}$ & 1.11 & (0.84, 1.39) & $c_{Feb24}$ & 7.12 & (3.58, 11.74) \\ 
  $\sigma^2_{\text{sur}}$ & 1.05 & (0.53, 1.73) & $c_{Mar23}$ & 3.58 & (2.57, 4.75) \\ 
  $\kappa_{\eta}$ & 0.85 & (0.69, 1.02) & $c_{Apr29}$ & 1.15 & (0.87, 1.48) \\ 
   &  &  & $\kappa_{\psi}$ & 0.92 & (0.31, 1.38) \\ 
\bottomrule
\end{tabular*}
\end{table*}

Posterior median estimates and 95\% HPD intervals for model parameters of interest are presented in Table~\ref{tab:real:par}. The analysis reveals no significant association between sea surface temperature and zooplankton abundance, nor between bathymetry and whale abundance. Importantly, we observe a significantly positive coefficient connecting zooplankton abundance and whale intensity. This suggests that areas with higher concentrations of zooplankton serve as favorable feeding grounds for whales. Fluctuations in zooplankton populations may directly influence whale distribution patterns and behaviors. These findings emphasize the importance of monitoring zooplankton dynamics in marine ecosystems to better understand and predict whale habitat use and overall ecosystem health. The surface tow underestimates the actual organism abundance. The log surface measurements are shifted by approximately 1.8 units when compared to the latent log zooplankton abundance. The average call rates per hour per whale are significantly different for the three days; we estimate 7 calls per hour per whale on February 24, 3.52 calls per hour per whale on March 23, and 1.15 calls per hour per whale on April 29. The estimates of the GP variance components imply that there is unexplained spatial variability for the spatial distributions of both zooplankton and whales.

\begin{figure}[tb]
    \centering
    \includegraphics[width = 0.78\textwidth]{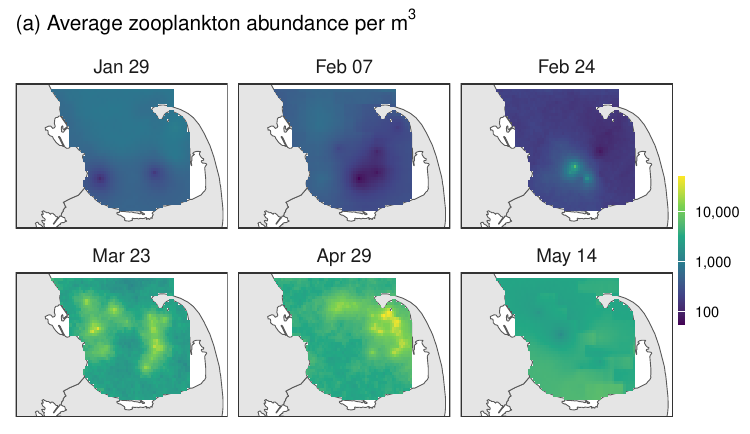}
    \includegraphics[width = 0.78\textwidth]{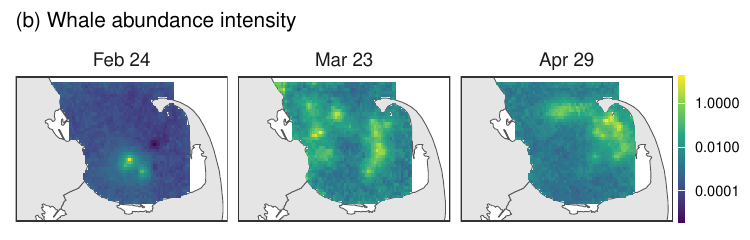}
    \caption{(a) Posterior mean estimate of the average zooplankton abundance surface per $m^3$ for each day. (b) Posterior mean estimate of the whale intensity surface for each day.}
    \label{fig:real:sur}
\end{figure}

Figure~\ref{fig:real:sur} (a) and (b) display the posterior mean estimates of the average zooplankton abundance surface and whale intensity surface, respectively, for each day. We observe smoother zooplankton surfaces for January 29, February 7, and May 14. This can be attributed to the fact that whale data were absent on those days, and relying solely on zooplankton data provides less information with respect to their spatial distribution. 
The spatial patterns of zooplankton and whale abundance exhibit similar trends across CCB on the shared observation dates of February 24, March 23, and April 29. This implies that the spatial distribution of zooplankton plays a consequential role in determining the spatial distribution of whales compared to other factors. It suggests a strong trophic linkage between prey and NARW, where whale movement and habitat use are closely tied to the availability and location of their primary food source. This dependence indicates that changes in zooplankton distribution, whether due to environmental shifts or anthropogenic influences, could directly impact whale foraging behavior, migration patterns, and population dynamics. Understanding this relationship is crucial for predicting whale responses to ecosystem changes and for informing conservation and management strategies aimed at protecting both zooplankton and whale populations.


\begin{table*}[tb]
\caption{Estimated posterior medians and 95\% HPD intervals for average zooplankton abundance per $m^3$ across CCB and for lower and upper bounds of total whale abundance for each day. \label{tab:real:abundance}}
\tabcolsep=0pt
\begin{tabular*}{\textwidth}{@{\extracolsep{\fill}}l rr rr rr @{\extracolsep{\fill}}}
\toprule%
\multirow{2}{*}{Date} & \multicolumn{2}{c}{\makecell{Zooplankton\\abundance}} & \multicolumn{2}{c}{\makecell{Lower bound\\of whale abundance}} & \multicolumn{2}{c}{\makecell{Upper bound\\of whale abundance}}\\
\cmidrule{2-3} \cmidrule{4-5}  \cmidrule{6-7}%
  & Median & 95\% HPD & Median & 95\% HPD & Median & 95\% HPD \\ 
  \midrule
  Jan 29 & 562 & (190, 1262) & - & -  & - & - \\ 
  Feb 07 & 314 & (81, 771) & - & -  & - & - \\ 
  Feb 24 & 226 & (90, 439) & 16 & (8, 27) & 16 & (8, 27) \\ 
  Mar 23 & 3934 & (2191, 6287) & 70 & (43, 105) & 133 & (91, 180) \\ 
  Apr 29 & 5153 & (1450, 12778) & 44 & (28, 63) & 131 & (98, 167) \\ 
  May 14 & 3341 & (1248, 6882) & - & -  & - & - \\ 
\bottomrule
\end{tabular*}
\end{table*}

Table~\ref{tab:real:abundance} presents the posterior median estimates and 95\% HPD intervals for abundance of zooplankton and whales for each day. Zooplankton abundance from March to May is estimated to be relatively higher than the levels estimated for January and February. The posterior median estimate of total whale abundance is only 16 on February 24, while the estimates for March 23 and April 29 are much higher, with lower and upper bounds of 70 and 133 for March 23, and 44 and 131 for April 29. On February 24, aerial behavioral observations were made for three individuals, none of whom were observed feeding. On March 23, there were behavioral observations for 45 NARWs, 15 of which were feeding. On April 29, there were 72 behavioral observations, all of which were of feeding whales. Given the estimated call rates of 7.12 on February 24, 3.58 on March 23, and 1.15 on April 29, these findings are consistent with anatomical evidence indicating that feeding whales do not make calls \citep{parks2011sound}.

\begin{figure}[tb]
    \centering
    \includegraphics[width=0.9\textwidth]{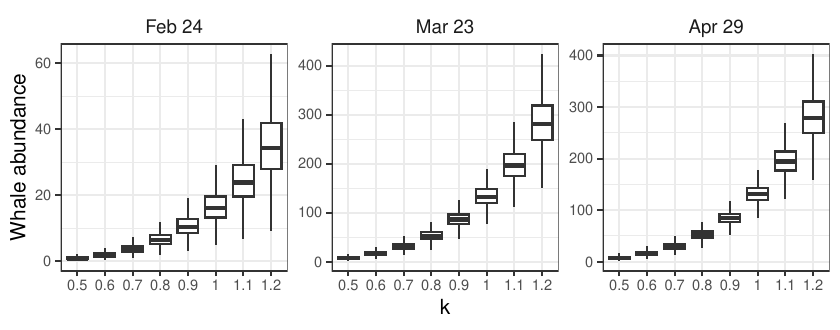}
    \caption{Box plots of posterior predictive samples for whale abundance given posterior samples for zooplankton abundance scaled by a factor of $k$ for each day.}
    \label{fig:real:zoopwhale}
\end{figure}

Lastly, given the complexity of our model specification (Figure~\ref{fig:model}), we offer more clarity to the NARW-prey relationship which the fitted model provides. In particular, we provide illustrative quantitative illumination of the effect of increasing zooplankton concentration on expected whale abundance for each day. We generate posterior predictive samples for whale abundance given posterior samples for zooplankton abundance scaled by a factor of $k \in \{0.5, 0.6, 0.7, 0.8, 0.9,$ $1.0, 1.1,1.2\}$. Figure~\ref{fig:real:zoopwhale} shows box plots of the posterior predictive samples for whale abundance for a set of $k$ values for each day. When zooplankton abundance is reduced by a half throughout CCB (i.e., $k$ = 0.5), the posterior predictive median estimate of whale abundance is approximately 1 on February 24, 8 on March 23, and 7 on April 29. When zooplankton abundance is increased by 20\% across CCB (i.e., $k$ = 1.2), the posterior predictive median estimate of whale abundance is approximately 34 for February 24, 281 for March 23, and 279 for April 29. These findings reveal the strong non-linear relationship between zooplankton abundance and whale populations in CCB. They further suggest that relatively small changes in zooplankton populations can lead to significant shifts in whale populations, particularly during key seasonal periods.

\section{Discussion}
\label{sec:discussion}

We have presented the first formal stochastic modeling investigation of the joint spatiotemporal relationship between a whale species (North Atlantic right whales) and their prey, zooplankton. In particular, employing two sources of prey data and two sources of whale data, we have developed a complex multi-level fusion model to address this challenge. Using Cape Cod Bay, MA as our sampling region, we have fit and compared two models within a Bayesian framework and have offered a simulation study to demonstrate that we can learn about the true joint relationship given adequate spatial coverage of data collection over CCB. Despite the limited spatial coverage and temporal alignment of our data, we have demonstrated that we can still fit the proposed model and obtain better inference about zooplankton abundance and whale abundance than modeling either individually. Further, our conditional times marginal specification enables the estimation and uncertainty quantification of whale abundance given zooplankton abundance. 
This study has the potential to support conservation efforts by shedding light on the spatial and temporal relationships between NARWs and their prey. Improved knowledge of these dynamics can inform strategic conservation efforts and policy decisions to preserve this endangered species amid changing ocean conditions.

Despite this being one of the best case studies in the world to model the relationship between a large whale species and its prey, there are limitations to our work. First, each of the data sources provides challenges to our modeling. We have a small number of sampled zooplankton sites resulting in relatively sparse coverage of CCB. 
In addition, the zooplankton data are comprised of several species.  Consideration of zooplankton dynamics throughout the season \citep{hudakNorthAtlanticRight2023} as well as how right whales respond could influence inference regarding their abundance.
With respect to NARWs, as discussed in \cite{schliep2023assessing}, the acoustic data is in the form of whale calls received at the fixed locations of the hydrophones rather than the locations of the whales. We chose to model the acoustic data in this fashion in part due to the clock drift inherent in these individual hydrophones, which limits localization \citep{palmerAccountingLombardEffect2022}. 
Second, there are a few important limitations to our model specification.  
The aerial distance sampling data is restricted to individuals that are ``available'' from above and we assume this availability is constant across space and the same for all individuals. Calling rates have been shown to exhibit diel behavioral variability according to activity patterns of individual, but the fusion limits us to the window of aerial data collection over which we assume calling rates are constant.  
Last, in modeling whale abundance, there may be other potential covariates not considered here, which can influence the spatial distribution and abundance of whales in CCB at a given time.  
Our approach provides a general framework for modeling this marine relationship and can be adapted to changes in the data sources. 


Future work entails examining roughly 90 potential days across several years where we have coincidence of sampling of all 4 data sources. This modeling of high dimensional data will be computationally demanding.  We will also consider localization of whale calls, which can be extracted from the hydrophone data through a very time consuming process. Here we modeled total zooplankton, but we also plan to explore the effect of how the abundance of different zooplankton species over the course of the season  \citep{hudakNorthAtlanticRight2023} may impact the joint relationship.

\section*{Funding sources and acknowledgments}
Aerial survey and zooplankton data for this study were collected under NOAA Scientific Permit No. 14603.
CCS acknowledges funding from NOAA Fisheries, Massachusetts DMF, Massachusetts Environmental Trust, and private donors. This work was supported in part by funding from NOAA Fisheries under award NA20NMF0080246, SERDP award RC20-1097, and US Office of Naval Research Awards N000142312562 and N000142412501.

\bibliographystyle{apalike}
\bibliography{refs}

\end{document}


\def\spacingset#1{\renewcommand{\baselinestretch}%
{#1}\small\normalsize} \spacingset{1}


\if1\blind
{
  \title{\bf Supplementary Material for ``Joint Spatiotemporal Modeling of Zooplankton and Whale Abundance in a Dynamic Marine Environment''}
  \author{Bokgyeong Kang, Erin M. Schliep, Alan E. Gelfand,\\ Christopher W. Clark, Christine A. Hudak, Charles A. Mayo,\\ Ryan Schosberg, Tina M. Yack, and Robert S. Schick
  }
  \maketitle
} \fi

\if0\blind
{
  \bigskip
  \bigskip
  \bigskip
  \begin{center}
    {\LARGE\bf Supplementary Material for ``??''}
\end{center}
  \medskip
} \fi

\spacingset{1.9} 



\section{Further details on the data sources}
\label{sup:sec:data}

In this section we provide additional details on the zooplankton observations and covariates used to describe latent zooplankton abundance and true whale intensity surfaces. 

\begin{figure}[tb]
    \centering
    \includegraphics[width = 4.5in]{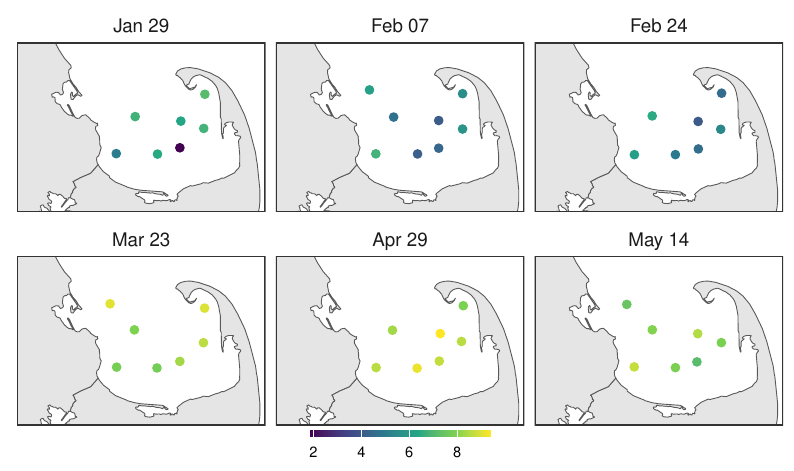}
    \caption{Oblique tow observations for each of the zooplankton observation days.}
    \label{sup:fig:obl}
\end{figure}

\begin{figure}[tb]
    \centering
    \includegraphics[width = 4.5in]{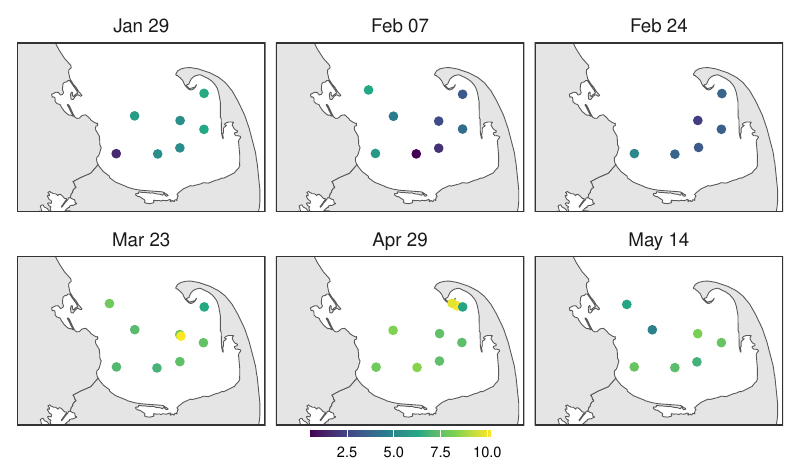}
    \caption{Surface tow observations for each of the zooplankton observation days.}
    \label{sup:fig:sur}
\end{figure}

Figure \ref{sup:fig:obl} shows the observed number of zooplankton organisms per $m^3$ at locations where the oblique tow was conducted for each of the zooplankton observation days. The surface tow observations are depicted in Figure \ref{sup:fig:sur}. Zooplankton abundance exhibits both temporal and spatial variability across CCB. In the early observation periods, zooplankton concentrations are generally lower, with isolated areas of higher density. As the season advances, zooplankton density increases, with some regions showing more substantial growth.

\begin{figure}[tb]
    \centering
    \includegraphics[width = 5.3in]{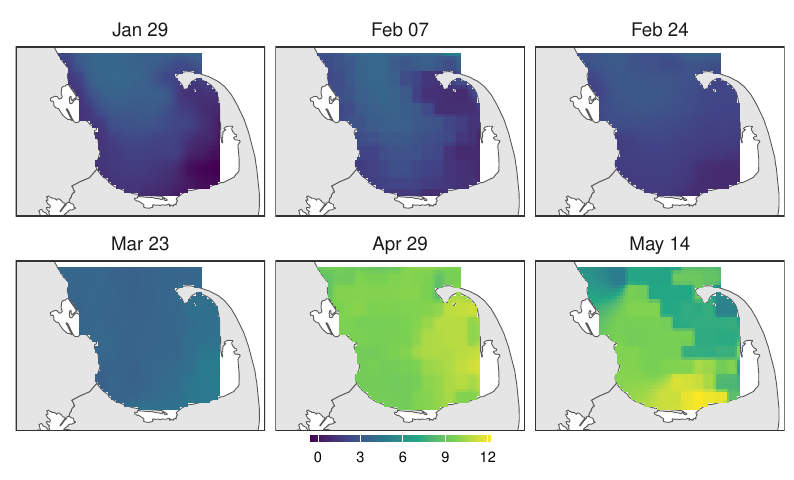}
    \caption{Kriged surface of sea surface temperature throughout CCB for each of the zooplankton observation days.}
    \label{sup:fig:temp}
\end{figure}

\begin{figure}[tb]
    \centering
    \includegraphics[width = 1.8in]{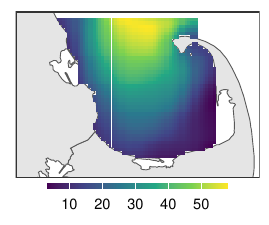}
    \caption{Bathymetry, reported in meters, throughout CCB.}
    \label{sup:fig:bath}
\end{figure}

We include sea surface temperature (SST) as a covariate in modeling the log-transformed latent zooplankton abundance surfaces, denoted as $Z{t_i}(\bs)$. SST data were sampled from the MODIS-Aqua OceanData satellite using Google Earth Engine. Owing to the presence of clouds in this area at this time of year, we extracted a 7-day moving average across CCB for each day in our analysis. The resulting geoTIFF was at a 1km resolution. Despite this moving average there were still some pixels with missing values.
Kriging, a geostatistical interpolation method, is employed to predict sea surface temperature at each of the grid cell centroids, and the resulting interpolated surfaces are shown in Figure~\ref{sup:fig:temp}. At the beginning of the observation period, cooler temperatures are predominant across the bay, especially in the northern regions. As the season progresses, a clear warming trend is observed, with higher temperatures increasingly spreading throughout CCB, particularly in the southern areas.

We incorporate bathymetry as a covariate in the LGCP model for the true whale intensity surface $\lambda_{t_i}(\bs)$. The bathymetry data were acquired from the SRTM Global Bathymetry and Topography dataset (\url{https://doi.org/10.5069/G92R3PT9}). Figure~\ref{sup:fig:bath} illustrates the bathymetry throughout CCB, highlighting a pronounced north-to-south gradient, as well as shallower depths near the coastline. All covariates were scaled prior to being included in the models.

\section{Additional simulation results}
\label{sup:sec:sim}

Due to space constraints, the main manuscript presents simulation results solely for day $t_3$. This section provides additional simulation results for the remaining days.



\begin{table*}[tb]
\caption{True values, posterior median estimates, and 95\% HPD intervals of average zooplankton abundance per $m^3$ across $\mathcal{D}$ (top) and total whale abundance (bottom) for Models (i) and (ii) fitted to their respective simulated data under real sampling designs of zooplankton and whale. \label{sup:tab:sim:abun12}}
\tabcolsep=0pt
\begin{tabular*}{\textwidth}{@{\extracolsep{\fill}}ll rrr rrr@{\extracolsep{\fill}}}
  \toprule%
  \multirow{2}{*}{Component} & \multirow{2}{*}{Day} & \multicolumn{3}{@{}c@{}}{Model (i)} & \multicolumn{3}{c}{Model (ii)}\\
  \cmidrule{3-5} \cmidrule{6-8}%
   & & Truth & Median & 95\% HPD & Truth & Median & 95\% HPD \\ 
  \midrule
  Zooplankton & $t_1$ & 313 & 212 & (55, 507) & 666 & 365 & (168, 640) \\ 
   & $t_2$ & 245 & 213 & (75, 473) & 447 & 349 & (176, 594) \\ 
   & $t_3$ & 146 & 178 & (62, 372) & 376 & 289 & (121, 501) \\ 
   & $t_4$ & 315 & 357 & (123, 776) & 320 & 319 & (164, 512) \\ 
   & $t_5$ & 682 & 839 & (310, 1541) & 546 & 542 & (199, 1118) \\ 
   & $t_6$ & 329 & 810 & (212, 2252) & 533 & 502 & (219, 939) \\ 
   \midrule
   Whale & $t_3$ & 30 & 19 & (7, 40) & 80 & 97 & (51, 161) \\ 
   & $t_4$ & 60 & 58 & (33, 92) & 50 & 53 & (30, 82) \\ 
   & $t_5$ & 241 & 269 & (218, 326) & 94 & 79 & (49, 118) \\ 
  \bottomrule
\end{tabular*}
\end{table*}

The top portion of Table~\ref{sup:tab:sim:abun12} displays the true values, posterior median estimates, and 95\% HPD intervals of average zooplankton abundance per $m^3$ across $\mathcal{D}$ for all zooplankton observation days for both models. For Model (i), the 95\% HPD intervals include the latent zooplankton abundance for each observation day. Except for $t_6$, the median estimates closely correspond to the latent values. On day $t_6$, where only zooplankton data are available and whale data are absent, the abundance is noticeably overestimated. For Model (ii), the median estimates are closely aligned with the latent zooplankton abundance, and the 95\% HPD intervals include the latent zooplankton abundance for all days except $t_1$. For day $t_1$, only zooplankton observations are available, with no accompanying whale data. This indicates that relying exclusively on zooplankton observations may result in overestimation or underestimation of the latent zooplankton abundance.

The lower part of Table~\ref{sup:tab:sim:abun12} shows the true values, posterior median estimates, and 95\% HPD intervals of total whale abundance for each of the joint observation days for Models (i) and (ii). Both models produce median estimates that are similar to the true values, with the 95\% HPD intervals covering the truth on all observation days.

\section{Further analysis results for the 2011 CCB data}
\label{sup:sec:real}

\subsection{Detection probabilities}
\label{sup:sec:real:detect}

\begin{figure}[tb]
    \centering
    \includegraphics[width = 0.85\textwidth]{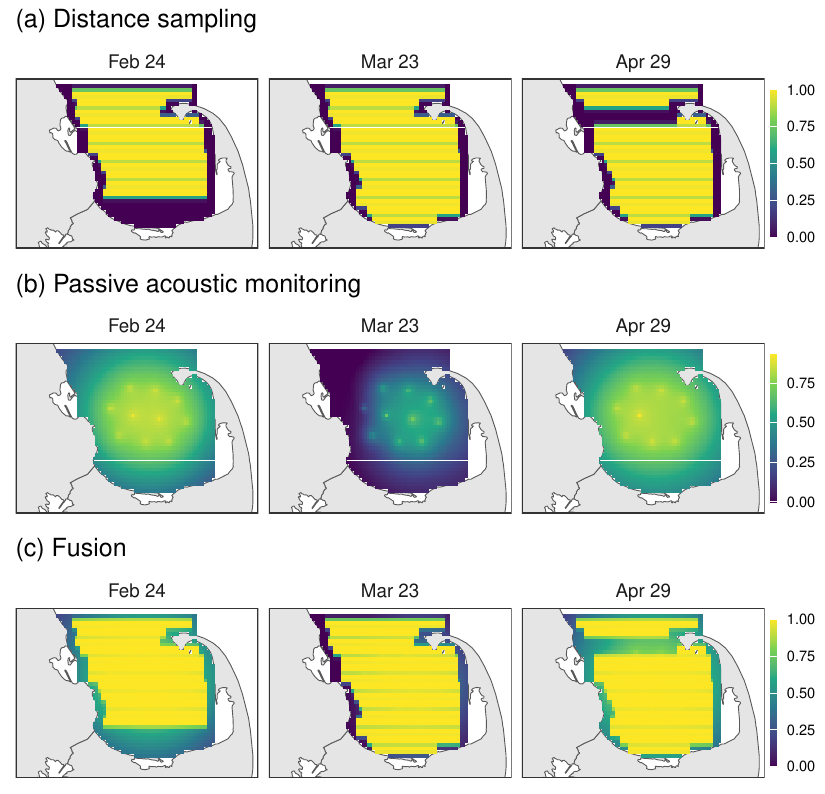}
    \caption{Detection probabilities for distance sampling, passive acoustic monitoring, or both given that the individual is on the surface and calling.}
    \label{fig:detectProb}
\end{figure}

We computed the detection probabilities across CCB for each whale observation day for distance sampling, which is conditional on a whale being at the surface at each location. These probability surfaces are shown in Figure~\ref{fig:detectProb} (a). The aerial detection probability is high near the transects. On February 24 and April 29, days of incomplete coverage, the detection probability is zero in the regions where the aerial survey was not conducted.

Figure~\ref{fig:detectProb} (b) displays the detection probabilities across the region for each whale observation day for passive acoustic monitoring, which is conditional on a whale producing a call. These detection probabilities were calculated using MARU- and day-specific median ambient noise levels, which ranged from 101.6 to 109.0 dB re:1µPa on February 24, 105.8 to 123.3 dB on March 23, and 100.6 to 107.1 dB on April 29. The higher noise levels on March 23 result in lower detection probabilities across the region on that day.

The detection probabilities resulting from the data fusion are shown in Figure~\ref{fig:detectProb}(c). The fusion improves detection probabilities in areas not covered by aerial surveys.

\subsection{Deviation from a standard distance sampling protocol}
\label{sup:sec:real:distsamp}

\begin{figure}[tb]
    \centering
    \includegraphics[width=0.98\linewidth]{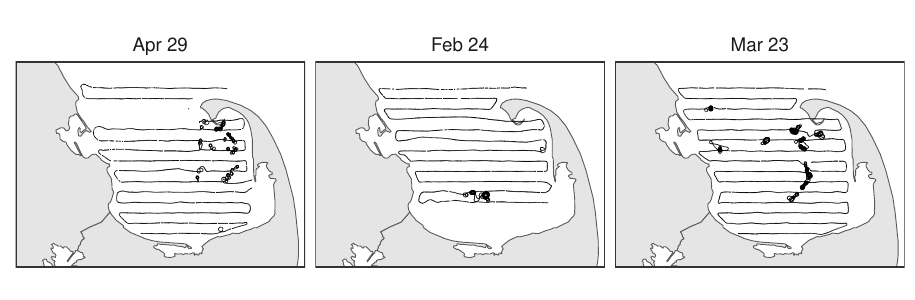}
    \caption{Actual flight path of the plane during the aerial survey on each observation day}
    \label{sup:fig:trajectory}
\end{figure}

Based on the standard protocol of distance sampling, observers follow straight lines (i.e., transects) and record the perpendicular distances from the line to each detected individual or group of animals. However, the aerial surveys conducted in CCB do not adhere to the strict distance sampling protocol. Figure~\ref{sup:fig:trajectory} depicts the actual flight path of the plane during the aerial survey on each observation day. When observers spot an individual or a group of animals, they approach and circle above the location and record the number of observed individuals (i.e., multiple whales can be recorded at the same location). This could result in detecting more individuals than anticipated under a strict distance sampling protocol.

\subsection{Inference results associated with the lower bound estimation for total whale abundance}
\label{sup:sec:real:lower}

As mentioned in the manuscript, the aerial surveys conducted in CCB did not strictly follow the distance sampling protocol. When observers detected an individual or group of animals, they approached and circled above the location, recording the number of individuals observed, which could result in multiple whales being recorded at the same location. This may lead to an overestimation of abundance compared to a strict distance sampling protocol. 
In the manuscript, we assume that the surveys adhered to the strict distance sampling protocol and included all recorded whale sightings, resulting in an \emph{upper} bound estimate of whale abundance. In contrast, here we assume that only one individual was recorded per location, regardless of the actual number of whales sighted. This approach provides a \emph{lower} bound for total whale abundance estimation.

\begin{table*}[tb]
\caption{Posterior median estimate and 95\% HPD interval for the loglikelihood function as well as CRPS for whale abundance detected by distance sampling for each day. \label{sup:tab:real:compLower}}
\tabcolsep=0pt
\begin{tabular*}{\textwidth}{@{\extracolsep{\fill}}l rr rrr@{\extracolsep{\fill}}}
\toprule%
\multirow{2}{*}{Model} & \multicolumn{2}{c}{Loglikelihood} & \multicolumn{3}{c}{CRPS}\\
\cmidrule{2-3} \cmidrule{4-6}%
  & Median & 95\% HPD & Feb 24 & Mar 23 & Apr 29 \\
  \midrule
  (i) & 10775 & (10753, 10799) & 2.17 & 1.56 & 1.14 \\ 
  (ii) & 10774 & (10750, 10796) & 1.89 & 1.65 & 1.20 \\ 
\bottomrule
\end{tabular*}
\end{table*}

For this dataset, we compare Models (i) and (ii) using the continuous ranked probability score (CRPS) for the number of whales detected via aerial distance sampling, as well as the posterior distributions of the loglikelihood function. The results, shown in Table~\ref{sup:tab:real:compLower}, indicate that the posterior median estimates of the loglikelihood function are comparable between the two models. Additionally, Model (i) produces CRPS values similar to those of Model (ii). Overall, these metrics suggest that both models offer comparable performance in fitting our dataset. Given that Model (i) is simpler while providing equivalent performance to Model (ii), we base further model inference on Model (i).

\begin{table*}[tb]
\centering
\caption{Estimated posterior medians and 95\% HPD intervals for model parameters. \label{sup:tab:real:parLower}}
\tabcolsep=0pt
\begin{tabular*}{\textwidth}{@{\extracolsep{\fill}}lrr lrr @{\extracolsep{\fill}}}
  \toprule
  \multicolumn{3}{c}{Zooplankton} & \multicolumn{3}{c}{Whale}\\
  \cmidrule{1-3} \cmidrule{4-6}
Parameter & Median & 95\% HPD & Parameter & Median & 95\% HPD \\ 
  \midrule
$\alpha_{temp}$ & 0.42 & (-0.93, 1.60) & $\beta_{bath}$ & 0.43 & (-0.64, 1.70) \\ 
  $\lambda^{sur}_{0}$ & -1.80 & (-3.57, -0.12) & $\beta_{zoop}$ & 4.44 & (3.48, 5.52) \\ 
  $\lambda^{sur}_{1}$ & 1.10 & (0.87, 1.35) & $c_{Feb 24}$ & 7.04 & (3.57, 11.72) \\ 
  $\sigma^2_{sur}$ & 1.00 & (0.52, 1.65) & $c_{Mar 23}$ & 6.13 & (4.07, 8.87) \\ 
  $\kappa_{\eta}$ & 0.68 & (0.53, 1.24) & $c_{Apr 29}$ & 3.25 & (2.11, 4.74) \\ 
  - & - & - & $\kappa_{\psi}$ & 0.82 & (0.24, 2.09) \\  
   \bottomrule
\end{tabular*}
\end{table*}

Posterior median estimates and 95\% HPD intervals for the model parameters of interest are presented in Table~\ref{sup:tab:real:parLower}. The inference results are largely consistent with those from the upper bound estimation in the manuscript, with the exception of the call rates on March 23 and April 29. For these dates, the call rates are estimated to be higher, likely due to fewer whales being observed via distance sampling while the number of detected calls remained the same.

\begin{figure}[tb]
    \centering
    \includegraphics[width = 0.8\textwidth]{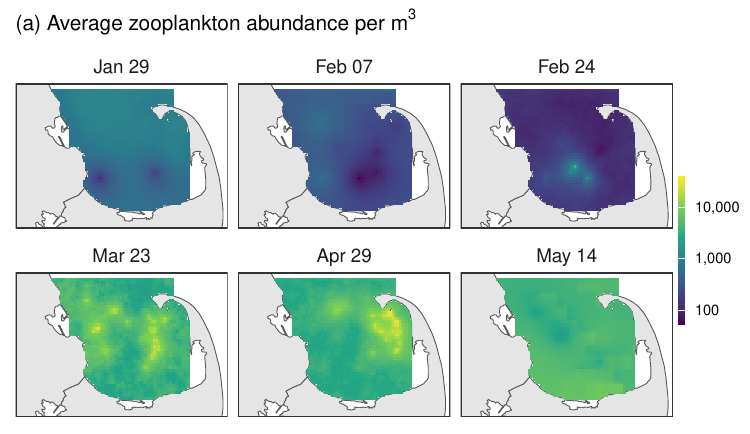}
    \includegraphics[width = 0.8\textwidth]{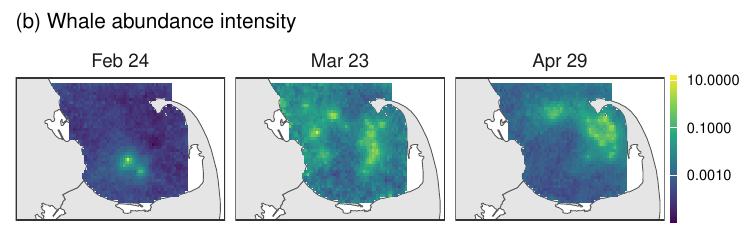}
    \caption{(a) Posterior mean estimate of the average zooplankton abundance surface per $m^3$ for each day. (b) Posterior mean estimate of the whale intensity surface for each day.}
    \label{sup:fig:real:surLower}
\end{figure}

Figure~\ref{sup:fig:real:surLower} (a) and (b) show the posterior mean estimates of the average zooplankton abundance surface and the whale intensity surface, respectively, for each day. The spatial patterns are very similar to those observed in the upper bound estimation.

\begin{table*}[tb]
\centering
\caption{Estimated posterior medians and 95\% HPD intervals for average zooplankton abundance per $m^3$ across CCB for each observation day. \label{sup:tab:real:abundanceZoopLower}}
\tabcolsep=0pt
\begin{tabular*}{\textwidth}{@{\extracolsep{\fill}}lrr @{\extracolsep{\fill}}}
\toprule%
Date & Median & 95\% HPD\\ 
  \midrule
  Jan 29 & 674 & (204, 1504) \\ 
  Feb 07 & 264 & (96, 552) \\ 
  Feb 24 & 158 & (83, 265) \\ 
  Mar 23 & 4637 & (2466, 7565) \\ 
  Apr 29 & 4570 & (2483, 7562) \\ 
  May 14 & 3873 & (1370, 8170) \\ 
\bottomrule
\end{tabular*}
\end{table*}

Table~\ref{sup:tab:real:abundanceZoopLower} provides the posterior median estimates and 95\% HPD intervals for the average zooplankton density (per $m^3$) across CCB for each day. The results are comparable to those obtained from the upper bound analysis in the manuscript.

\begin{figure}[tb]
    \centering
    \includegraphics[width=0.9\textwidth]{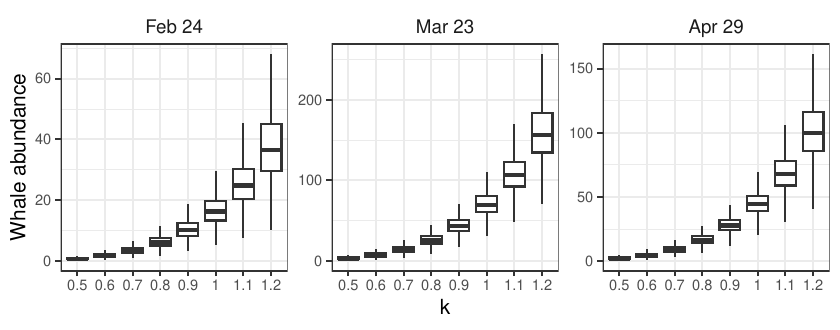}
    \caption{Box plots of posterior predictive samples for whale abundance given posterior samples for zooplankton abundance scaled by a factor of $k$ for each day.}
    \label{sup:fig:real:zoopwhaleLower}
\end{figure}

Lastly, we provide quantitative illumination of the effect of increasing zooplankton concentration on expected whale abundance for each day. We generate posterior predictive samples for whale abundance given posterior samples for zooplankton abundance scaled by a factor of $k \in \{0.5, 0.6, 0.7, 0.8, 0.9,$ $1.0, 1.1,1.2\}$. Figure~\ref{sup:fig:real:zoopwhaleLower} illustrates the box plots of the posterior predictive samples for whale abundance for an array of $k$ values for each day. When zooplankton abundance is reduced by a half throughout CCB (i.e., when $k$ = 0.5), the posterior predictive median estimate of whale abundance is approximately 1 on February 24, 3 on March 23, and 2 on April 29. When zooplankton abundance is increased by 20\% across CCB (i.e., when $k$ = 1.2), the posterior predictive median estimate of whale abundance is approximately 37 for February 24, 157 for March 23, and 100 for April 29.
